\documentstyle[12pt,aaspp4]{article} 

\def\kms{km~s$^{-1}$}

\def\msun{$M_{\odot}$}
\def\te{$T_e$}
\def\ne{$n_e$}
\def\dhii{$d_{\sc HII}$}
\def\cmthree{cm$^{-3}$}
\def\cm3{cm$^{-3}$}
\def\cmsix{cm$^{-6}$}

\def\a220{Arp~220~}
\def\mhii{M$_{\sc HII}$~}
\def\nlyc{N$_{\sc Lyc}$~}
\def\nhiilos{N$_{los}^{\sc HII}$}
\def\lbol{L$_{bol}$}
\def\lsun{L$_\odot$}

\def\sc{\scriptstyle}

\def\deg{$^{\circ}$}

\def\a{$\alpha$}
\def\g{$\gamma~$}

\begin{document}


\title{Starburst in the Ultra-luminous Galaxy Arp 220 - Constraints from
         Observations of Radio Recombination Lines and Continuum}

\bigskip

\author{K.R. Anantharamaiah\altaffilmark{1,2}, 
        F. Viallefond\altaffilmark{3},  
        Niruj R. Mohan\altaffilmark{1,2,4}, 
        W.M. Goss\altaffilmark{1} \& 
        J.H. Zhao\altaffilmark{5} }

\bigskip

\affil{\altaffilmark{1}National Radio Astronomy Observatory, Socorro, 
       NM 87801, USA}
\affil{\altaffilmark{2} Raman Research Institute,  Bangalore 560 080, 
       India}
\affil{\altaffilmark{3} DEMIRM, Observatoire de Paris, F-75014 Paris, France}
\affil{\altaffilmark{4}Joint Astronomy Program, Indian Institute of Science,
       Bangalore 560 012, India}
\affil{\altaffilmark{5}Center for Astrophysics, Cambridge, MA 02138, USA}

\medskip

\centerline{email: anantha@rri.ernet.in, fviallef@maat.obspm.fr, 
           niruj@rri.ernet.in}
\centerline{mgoss@aoc.nrao.edu, jzhao@cfa.harvard.edu}

\bigskip

Short title: Ionized Gas in Arp 220



\begin{abstract}

We present observations of radio recombination lines (RRL) from the
starburst galaxy Arp 220 at 8.1 GHz (H92\a) and 1.4 GHz (H167\a\- and
H165\a) and at 84 GHz (H42\a), 96 GHz (H40\a) and 207 GHz (H31\a)
using the Very Large Array and the IRAM 30 m telescope,
respectively. RRLs were detected at all the frequencies except at 1.4
GHz where a sensitive upper limit was obtained. We also present
continuum flux measurements at these frequencies as well as at 327 MHz
made with the VLA. The continuum spectrum, which has a spectral index
$\alpha \sim -0.6$ ($S_\nu \propto \nu^{\alpha}$) between 5 and 10
GHz, shows a break near 1.5 GHz, a prominent turnover below 500 MHz
and a flatter spectral index above 50 GHz.

We show that a model with three components of ionized gas with
different densities and area covering factors can consistently explain
both RRL and continuum data.  The total mass of ionized gas in the
three components is $3.2\times 10^7$ \msun\- requiring $3\times 10^5$
O5 stars with a total Lyman continuum (Lyc) production rate \nlyc
$\sim 1.3\times 10^{55}$ photons s$^{-1}$.  The ratios of the expected
to observed Br\a\- and Br$\gamma$ fluxes implies a dust extinction
$A_V \sim 45$ magnitudes. The derived Lyc photon production rate
implies a continuous star formation rate (SFR) averaged over the life
time of OB stars of $\sim$ 240 \msun\- $yr^{-1}$.  The Lyc photon
production rate of $\sim$ 3\% associated with the high density HII
regions implies similar SFR at recent epochs ($t<10^5$ yrs).  An
alternative model of high density gas, which cannot be excluded on the
basis of the available data, predicts ten times higher SFR at recent
epochs. If confirmed, this model implies that star formation in Arp
220 consists of multiple starbursts of very high SFR (few $\times
10^3$ \msun\- yr$^{-1}$) and short durations ($\sim 10^5$ yrs).

The similarity of IR-excess, $L_{IR}/L_{Ly\alpha}\sim 24$, in Arp 220
to values observed in starburst galaxies shows that most of the high
luminosity of Arp 220 is due to the on-going starburst, rather than
produced by a hidden AGN. A comparison of the IR-excesses in Arp 220,
the Galaxy and M33 indicates that the starburst in Arp 220 has an IMF
which is similar to that in normal galaxies and has a duration longer
than $10^7$ yrs. If there was no infall of gas during this period,
then the star formation efficiency (SFE) in Arp 220 is $\sim$50\%. The
high SFR and SFE in Arp 220 is consistent with their known dependences
on mass and density of gas in star forming regions of normal galaxies.

\end{abstract}


\section{Introduction}

Arp 220 is the nearest ($d$ = 73 Mpc for $H_0$ = 75 \kms Mpc$^{-1}$)
example of an ultra-luminous
infrared galaxy (ULIRG). ULIRGs are a class of galaxies with enormous
infrared luminosities of $L_{IR} \ge 10^{12}L_\odot$,
believed to have formed through mergers of two gas-rich galaxies
(Sanders et al 1988).  What powers the high luminosity of ULIRGs is
not well understood, although there is considerable evidence that intense
starburst which may be triggered by the merger process may play a
dominant role (Sanders and Mirabel 1996, G\"{e}nzel et al 1998). There
is also evidence in many ULIRGs for a dust-enshrouded AGN which could
contribute significantly to or in some cases dominate the observed
high luminosity (G\"{e}nzel et al 1998). In the case of Arp 220, there is
clear evidence for a merger in the form of extended tidal tails
observed in the optical band (Arp 1966) and the presence of a double
nucleus in the radio and NIR bands with a separation of $\sim 1''$
(370 pc)(Norris 1985, Graham et al 1990). Prodigious amounts of
molecular gas ($M_{H_2} \sim 10^{10} M_\odot$; Scoville, Yun \& Bryant
1997, Downes \& Solomon 1998) are present in the central few hundred
parsecs.  Detection of CS emission (Solomon, Radford \& Downes 1990)
indicates that the molecular densities may be high ($n_{H_2}\sim 10^5$
\cmthree).  Recent high resolution observations reveal that the two
nuclei in Arp 220 consists of a number of compact radio sources -
possibly luminous radio supernovae (Smith et al 1998) - and a number
of luminous star clusters in the near IR (Scoville et al 1998). High
resolution observations of OH megamasers (Lonsdale et al 1998) reveal
multiple maser spots with complex spatial structure. These
observations have led to the suggestion that the nucleus of Arp 220 is
dominated by an intense, compact starburst phenomena rather than the
presence of a hidden AGN (Smith et al 1998, Downes \& Solomon
1998). The presence of AGNs in the two nuclei is, however, not ruled
out since large velocity gradients, $\sim 500$ \kms\- over $\sim 1''$
at different position angles across each nuclei (Sakamoto et al 1999)
have been observed, indicating high mass concentrations at the
location of both nuclei. Obscuration due to dust in the nuclear region
of Arp 220 is extremely high. Ratios of fine structure lines observed
with the ISO satellite (Lutz et al 1996, G\"{e}nzel et al 1998) indicate a
lower limit of $A_V \sim$ 45 mag. This high extinction seriously
hampers measurements even at IR wavelengths (see also Scoville et al
1998).

Radio recombination lines (RRLs) do not suffer from dust obscuration
and thus may potentially provide a powerful method of studying the
kinematics, spatial structure and physical properties of ionized gas
in the nuclear region of galaxies.  RRLs have been detected, to date,
in about 15 Galaxies (Shaver, Churchwell and Rots 1977, Seaquist and
Bell 1977, Puxley et al 1991, Anantharamaiah et al 1993, Zhao et al
1996, Phookun, Anantharamaiah and Goss, 1998).  Arp 220 is the
most distant among the galaxies that are detected in recombination lines
(Zhao et al 1996). The detection of an RRL in Arp 220 (Zhao et al
1996) opens up the possibility of studying the ionized gas in the
nuclear region.  Zhao et al (1996) presented VLA observations of the
H92\a\- line ($\nu_{rest}$ =8.309 GHz) with an angular resolution of
$\sim 4''$ and a velocity coverage of $\sim 900$ \kms.  This paper
presents new H92\a\- VLA observations with an angular resolution of
$\sim 1''$ and a velocity coverage of $\sim 1700$ \kms.  We also
present observations of millimeter wavelength RRLs H42\a\-
($\nu_{rest}$ =85.688 GHz), H40\a ($\nu_{rest}$ = 99.022 GHz) and
H31\a\- ($\nu_{rest}$ = 210.501 GHz) made using the IRAM 30 m radio
telescope in Pico Veleta, Spain. Additionally, we report a sensitive
upper limit to lower frequency RRLs H167\a\- ($\nu_{rest}$ = 1.40 GHz)
and H165\a ($\nu_{rest}$ = 1.45 GHz ) obtained using the VLA. These
observations together with the observed continuum spectrum are used to
constrain the physical properties and the amount of ionized gas in the
nuclear region of Arp~220. Since the Lyman continuum photons which
ionize the gas are generated by young O and B stars with a finite life
time, the derived properties of the ionized gas are  used to
constrain the properties of the starburst in Arp 220.

The larger velocity coverage used for the present H92\a\- observations
showed that the previous observation with half the bandwidth (Zhao et
al 1996) had underestimated both the width and the strength of the
H92\a\- line. The integrated line intensity is found to be more than a
factor of two larger than reported earlier and thus the deduced number
of Lyman continuum photons that are required to maintain the
ionization is also larger.  The millimeter wavelength lines are found
to be much stronger than  expected based on models for the
H92\a\- line. The observed strength of the millimeter wavelength RRLs
indicate the presence of a higher density ($n_e > 10^5$ \cmthree)
ionized gas component and leads to information about star formation rates at
recent epochs ($t < 10^5$ yrs).  The upper limits to the lower
frequency lines provide significant constraints on the amount of low
density ($n_e < 10^3$ \cmthree) ionized gas in the nuclear region of
Arp 220. A comparison of the predictions (based on RRLs) of the strengths
of NIR recombination lines (e.g. Br\a\- and Br\g lines) with
observations (e.g. Goldadder et al 1995, Sturm et al 1996) leads to a
revised estimate of the dust extinction.

This paper is organized as follows. In Section 2, observations with
the the VLA and the IRAM 30 m telescope are described and the results
are presented. Section 3 summarizes the available radio continuum
measurements of Arp 220 in the frequency range 150 MHz to 230 GHz
including new measurements at 327 MHz, 1.4 GHz, 8.3 GHz, 99 GHz and
206 GHz. In Section 3, we discuss the various possible models to
explain the observed RRLs and continuum in Arp 220 and arrive at a
three-component model, consistent with the observations.  Section
4  discusses the implications of the three-component model to several
aspects of the starburst in Arp 220.  We attempt to
answer the question of whether the deduced star formation rates in Arp
220 can power the observed high luminosity  without invoking
the presence of an obscured active nucleus. The paper is summarized in
Section 5.

\section{Observations, Data Reduction and Results}

\subsection{VLA observations and Results}

\subsubsection{H92\a\- line}

Observations of the H92\a\- line ($\nu_{rest}$ = 8309.38 MHz) were made
in Aug 1998 in the B configuration of the VLA.  The parameters of the
observations are given in Table 1. Phase, amplitude and frequency
response of the antennas were corrected using observations of the
calibrator source 1600+335 ($S_{3.5cm}\sim $ 2 Jy) every 30 minutes.
Standard procedures in AIPS were used for generating the continuum
image and the line cube. A self-calibration in both amplitude and
phase was performed on the continuum channel and the solutions were
applied to the spectral line data. For subtracting the continuum, two
channels, one on each end (channels 2 and 14), were used in the
procedure UVLIN (Cornwel, Uson and Haddad 1992).  During off-line data
processing, it became necessary to apply hanning smoothing which
resulted in a spectral resolution of 6.25 MHz (i.e. 230 km/s) with two
spectral points per resolution element. The synthesized beam with
natural weighting of the visibility function was $1.1''\times 0.9''$,
PA = 74\deg.
 
The continuum image of Arp220 is shown in Fig 1 as contours with the
image of the velocity-integrated H92\a\- line (moment 0) superposed in
grey scale. The two radio nuclei of Arp 220 (Norris et al 1988) which
are separated by about $1''$ are barely resolved. Fig 1 shows that the
recombination line emission is detected from both nuclei.  The peak of
the line emission is slightly offset to the north with respect to the
continuum peak.  The position angle of the elongated line emission is
tilted with respect to the line joining the two continuum peaks by
$\sim +7^\circ$. Fig 2 shows a contour image of the velocity-integrated
line emission (moment 0)  with velocity dispersion (moment2)
superposed in grey scale.  The recombination line is stronger and
wider near the western source and there appears to be a weak extension
of the ionized gas towards the north-east.

Fig 3  shows the integrated H92\a\- line profile (top frame)
along with line profiles near the eastern and western continuum peaks.
The line profile is wider on the western peak. The parameters of the
integrated line profile are given in Table 2.  The
spatially-integrated H92\a\- line has a peak flux density of 0.6
mJy/beam and a FWHM of 363 \kms.  Both these values are larger than
the values reported by Zhao et al (1996).  The integrated line flux is
8$\times 10^{-23}$ W m$^{-2}$ which is a factor of 2.3 larger than the
values reported by Zhao et al (1996). The integrated flux
was underestimated by Zhao et al (1996) since a narrower bandwidth
was used in their observation, thereby missing a portion of the line.

Because of the coarse velocity resolution (230 \kms) of the H92\a\-
data, detailed information about the kinematics is  limited. Fig 4a shows a
position-velocity diagram along the line joining the two continuum peaks
(P.A. = 98\deg). The velocity range is wider near the western peak
compared to the eastern peak.  After correcting for the instrumental
resolution, the full velocity range over the western peak is about 700
\kms .  On the eastern peak the velocity range is 440 \kms. The FWHMs
on the western and the eastern peaks are 500 \kms and 340 \kms
respectively.  Fig 4b shows a position-velocity diagram along the line
perpendicular to the line joining the two peaks.  The full velocity
range of emission in Fig 4b is about 770 \kms with a FWHM of 430
\kms. No velocity gradients are visible in Figs 4a and 4b.
 
\subsubsection{H167\a\-  and H165\a\- Lines}

The H167\a\- ($\nu_{obs}$ = 1399.37 MHz) and the H165\a\- ($\nu_{obs}$
= 1450.72 MHz) lines were observed simultaneously over a 7 hour period
in March 1998 in the A configuration of the VLA. Observational details
are summarized in Table 1.  Data were processed using standard
procedures in AIPS. With natural weighting of the visibilities, the
beam size is $1.8''\times 1.6''$. The peak continuum flux density was
215 mJy/beam and the integrated flux density is 312$\pm$3 mJy.  After
hanning smoothing, the resolution of each spectral channel was 390 kHz
($\sim 84$ \kms) and the rms noise was $\sim 130~ \mu$Jy/beam.
Neither line was detected.  For a line width of 350 \kms, we obtain a
$3\sigma$ upper limit to the line strength of 0.25 mJy.  The
parameters are listed in Table 2. As shown in the next section, the
upper limit to the RRLs near 1.4 GHz provide significant constraints
on the density of ionized gas in Arp 220.

\subsection{IRAM-30 m Observations  and Results}

Observations with the IRAM-30 m telescope in Pico Veleta, Spain were
carried out on April 19, 1996.  The pointing of the telescope was
adjusted from time to time; the rms pointing error was $\sim 3''$.
Three receivers were used simultaneously.  A 1.2 mm receiver connected
to a 512 MHz filter bank spectrometer was used to observe the
H31$\alpha$ line ($\nu_{rest}$ = 210.5018 GHz). The other two
receivers were used to observe the H40$\alpha$ ($\nu_{rest}$ = 99.0229
GHz) and the H42$\alpha$ ($\nu_{rest}$ = 85.6884 GHz) lines in the 3 mm
band.  The two 3 mm lines were observed using two subsets of the
autocorrelation spectrometer, each with 409 channels.  Instrumental
parameters are summarized in Table 3.  In the 1.2 mm band, the system
temperature ranged between 260 K and 285 K and in the 3 mm band, the
system temperatures were between 180 K and 200 K. The wobbler was used
to remove the sky emission and to determine the instrumental frequency
response.  Spectra were recorded every 30 seconds.

The data from the IRAM-30 m telescope were reduced using a special
program written by one of us (FV) to reduce single-dish spectral data
in the Groningen Image Processing System (GIPSY) environment (van der
Hulst et al 1992).  In this new program, a baseline subtracted final
spectrum as well as a continuum level are simultaneously determined
from the data base which consists of a series of spectra sampled every
30 s. The temporal information is used to perform statistics on the
data to measure the rms noise as a function of integration time and
frequency resolution. These statistics also provide the uncertainties
associated with the fitted baselines. These measured uncertainties
are used in the determination of error bars in the line and continuum
flux densities. In addition, the program also determines a statistical
rms noise level for each spectral channel separately, which helps in
judging the reality or otherwise of a spectral feature.  The details
of this procedure are described by Viallefond (2000, in preparation).

The final spectra obtained with the IRAM-30 m data are shown in Fig 5
and the parameters are listed in Table 4.  The quoted error bars are the
quadrature sums of the system noise and baseline uncertainties. To
obtain the H42\a\- spectrum shown in the bottom panel of Fig 5, linear
baselines were fitted over the velocity ranges 4600-5050 \kms and
5800-6200 \kms.  The spectrum has been smoothed to a resolution of 30
MHz and re-sampled to obtain 17 independent frequency channels out of
the original 409 channels. The H42\a\- recombination line extends from
5050 \kms to 5800 \kms.  The emission feature at the higher velocity
end of the spectrum (i.e $v >$ 6200 \kms) is probably the rotational
line of the molecule C$_3$H$_2$ ($\nu_{rest}$ = 85.339 GHz). The
continuum level is poorly determined from this data set and thus only
a $3\sigma$ upper limit to the continuum flux density is listed in Table 4.

The H40\a\- spectrum shown in the middle panel of Fig 5 was
determined by fitting linear baselines to channels outside the
velocity range 4950-5850 \kms. The spectrum has been smoothed to a
resolution of 30 MHz and re-sampled at 17 independent frequency
points.  The dashed lines in Fig 5 shows the rms noise across the
spectrum determined as described in Viallefond (2000, in preparation). The
H40\a\- line emission is quite similar to the H42\a\- emission in Fig
5. The consistency between the two lines confirms the reality of the
detections. This data set also provided a measurement of the continuum
flux density near 3 mm which is listed in Table 4

The observations of the H31\a\- line near 202.7 GHz have a serious
limitation.  The 512 MHz ($\sim$750 \kms) bandwidth used for the
observations was too small to cover the entire extent of the line
(FWZI $>$ 800 \kms).  To determine the continuum emission and the
spectral baseline level, we used the few channels at one end of the
spectrum where no line emission was expected.  Although the resulting
spectrum, shown as a solid line in the top panel of Fig 5, covers only
a part of the line emission, it is consistent in shape with the other
recombination lines in Fig 5.  The statistical significance of the
detection of the H31\a\- line is $> 10\sigma$ near the peak.  Table
2 lists the ratios of the strengths of various recombination lines
integrated over the partial velocity range 5220-5755 \kms over which
the H31\a\- line has been observed. The strength of the H31\a\- line
over the full extent of the line is estimated based on these ratios.

The millimeter wavelength recombination lines (H42\a, H40\a\- and H31\a)
detected with the IRAM-30 m telescope are much stronger than the
centimeter wavelength line (H92\a) detected with the VLA.  While the
peak line strength at $\lambda$ = 3.5 cm (H92\a) is 0.6 mJy, the
corresponding lines  at $\lambda$ = 3 mm (H40\a) and 1.2 mm
(H31\a) are an order of magnitude more intense: 15 mJy and 80
mJy, respectively.  An increase in the recombination line strength
at millimeter wavelengths was expected from high density models
discussed by Anantharamaiah et al (1993) and Zhao et al (1996).
However, the observed increase in the line strength is much
larger than expected on the basis of the models for the H92\a\- line.
The mm RRLs  H40\a\- and H42\a\- also appear to be broader than
the cm RRL H92\a\- line. The width of the mm RRLs are comparable
to that of CO lines (Downes and Solomon 1998).

\section{Radio Continuum Spectrum of Arp 220}

As a byproduct of the RRL measurements, continuum flux densities are
obtained at 1.4, 8.2, 97.2 and 206.7 GHz. The measured flux densities
are listed in Table 5 along with published values at  other
frequencies. The continuum flux density measurements around 1.5 mm are consistent
with the determination by Downes and Solomon (1998), but the value
near 3 mm is higher than that reported by the same authors.
The increase in the flux density above 200 GHz is due to contribution
by thermal dust emission.

In addition, we observed Arp 220 near 0.325 GHz in August 99 using the
A-configuration of the VLA using 3C286 as the flux calibrator. The
angular resolution was $\sim 6.5''$. Arp 220 is unresolved with this
beam. The integrated flux density of Arp 220 is 380$\pm$15 mJy.  A
flux density measurement at an even lower frequency of 0.15 GHz has
been reported by Sopp and Alexander (1991), which is also included in
Table 5.

The continuum spectrum in the frequency range 0.1 to 100 GHz is
plotted in Fig 6. The continuum spectrum is non-thermal in the
frequency range 3 to 30 GHz with a spectral index $\alpha\sim -0.6$
(where $S_\nu\propto \nu^\alpha$). The spectrum changes both below and
above this frequency range. A break in the spectrum is observed around
2 GHz with a change in the spectral index to $\sim -0.1$ between 1.6
GHz and 0.325 GHz. A further change in spectrum occurs below 300
MHz. The spectral index changes to $\sim +0.5$ and between 0.325 GHz
and 0.15 GHz.  These changes in the spectrum at lower frequencies are
complex and cannot be explained by a simple free-free absorbing
thermal screen or thermal gas mixed with the non-thermal component
(e.g.  Sopp \& Alexander 1991).  At the higher frequencies the change
in the spectrum is gradual. The spectral index appears to become
slightly flatter (i.e. $\alpha > -0.6$) above 22.5 GHz.  The solid
line in Fig 6 is a fit to the continuum spectrum based on a
3-component ionized gas model (next section) developed to explain the
observed recombination lines.

\section {Models for  RRL and Continuum emission}

Models for RRL emission from the nuclear region of external galaxies
have been discussed by Puxley et al (1991), Anantharamaiah et al
(1993), Zhao et al (1996, 1998) and Phookun et al (1998).  The main
constraints for these models are the integrated RRL strength at one or
more frequencies, the observed radio continuum spectrum and
geometrical considerations.  Two types of models have been considered:
one based on a uniform slab of ionized gas and the other based on a
collection of compact HII regions.  The observed non-thermal radio
continuum spectrum of the nuclear region, with spectral index $\alpha
\le -0.6$ (where $S_\nu \propto \nu^{-\alpha}$), imposes strong
constraints on the nature of the ionized gas that produces the
observed recombination lines from the same region.  If the models are
constrained by a single RRL measurement and the continuum spectrum,
then the derived physical parameters ($T_e, n_e, M_{HII}$ etc) are not
unique.  In a majority of the cases that were considered
(Anantharamaiah et al 1993, Zhao et al 1996, Phookun et al 1998), the
model with a collection of compact, high density HII regions is
favored. The uniform slab models produce an excess of thermal
continuum emission at centimeter wavelengths inconsistent with the
observed non-thermal spectrum.  However, in the case of Arp 220, Zhao
et al (1996) were able to fit both types of models; thus further
observations at higher and lower frequencies were required to choose
between the models. We consider below these models of RRL emission in
the light of the new measurements.

\subsection{Uniform Slab Model}

In the uniform slab model, the parameters are the electron temperature
(\te), the electron density (\ne) and the thickness of the slab ({\em
l}). For a given combination of \te\- and \ne, {\em l} is adjusted to
account for the observed integrated H92\a\- line strength. While
calculating the line strength, stimulated emission due to the
background non-thermal continuum, as well as internal stimulated
emission due to the thermal continuum emission from the ionized gas
are taken into account.  The relevant expressions for the computation
are given by Anantharamaiah et al (1993). The observed total continuum
emission at two frequencies, together with the computed thermal
emission from the ionized slab, is used to estimate the intrinsic
non-thermal emission and its spectral index.  Models in which the
thermal emission from the slab at 5 GHz exceeds a substantial
fraction ($\sim$ 30 -- 50\%) of the total continuum emission are
rejected since the resulting spectral index will not be consistent
with the observed non-thermal spectrum.  Finally, the expected
variation of line and continuum emissions as a function of frequency
are computed and compared with the observations.

Fig 7a shows the expected variation of integrated RRL strength as a
function of frequency for \te = 7500 K and three values of electron
density. Fig 7b shows the expected variation of continuum emission for
the corresponding models. In these models, the slab of ionized gas
is assumed to be in front of the non-thermal source. Results are
qualitatively similar if the ionized gas is mixed with the non-thermal
gas. The models are normalized to the H92\a\- line and
the continuum flux densities at 4.7 and 15 GHz.   Fig
7 shows that while the non-thermal continuum spectrum above 2 GHz and
the change in the spectrum near 1.6 GHz agree with  the models,
the flux densities at 0.15 and 0.325 GHz are not explained. In all
the models, free-free absorption is prominent below 1.5 GHz. No uniform
slab model can be found which both accounts for the H92\a\- line
and also produces a turnover in the continuum spectrum at a frequency
$<$ 1.5 GHz.  Furthermore, all the models that fit the
H92\a\- line predict a strength for RRLs  near 1.4 GHz line 
inconsistent with the upper limit. For other values of
\te\- (5000 K and 10000 K), the curves are similar to Figs 7a and 7b. No
models with $n_e > 10^3$ \cm3 could be fitted to the H92\a\- line. The
parameters of the models shown in Fig 7 are given in Table 6. In these
models, external stimulated emission (i.e. due to the background
non-thermal radiation) accounts for more than 75\% of the H92\a\- line
strength. No uniform slab model, at any density, could be fitted to
the higher frequency (H42\a, H40\a\- and H31a) lines. Since all the
uniform slab models are  inconsistent with low frequency RRL and
continuum data, we consider models with a collection of HII regions.

\subsection{Model with a Collection of HII regions}

In this model, the observed RRLs are thought to arise in a number of
compact, high density HII regions whose total volume filling factor in
the nuclear region is small ($< 10^{-4}$) (Puxley et al 1991,
Anantharamaiah et al 1993).  The low volume filling factor ensures
that the HII regions, regardless of their continuum opacities, have
only a small effect on the propagation of the non-thermal continuum
which originates in the nuclear region. Thus, the observed radio
continuum spectrum can be non-thermal even below the frequency at
which the HII regions themselves become optically thick.  The low
filling factor also implies that in these models there is little
external stimulated emission.  Since the HII regions are nearly
optically thick at centimeter wavelengths, they produce only a modest
amount of thermal continuum emission (typically $<$ 30\% of
S$_{obs}$).  The recombination line emission from these regions arises
mainly through internal stimulated emission due to the continuum
generated within the HII regions. If the density of the HII regions
are lower ($n_e < 10^3$ \cmthree) and the area covering factor $f_c$
(i.e. the fraction of the area of the non-thermal emitting region that
is covered by HII regions) is larger, then external stimulated
emission can become dominant at centimeter wavelengths.

For simplicity, all the HII regions are considered to be
characterized by the same combination of \te, \ne\-  and diameter
\dhii.  For a given combination of these parameters, the expected
integrated line flux density ($\int S_ld_\nu$) of a single HII region is
calculated using standard expressions (e.g. Anantharamaiah et al
1993).  The number of HII regions is then computed by dividing the
integrated flux density of one of the observed RRLs (e.g. H92\a) by
the expected strength from a single HII region. Since the volume
filling factor of the HII regions is small, the effect of shadowing of
one HII region by another is not significant. Constraints from the
observed continuum flux densities at various frequencies and from
several geometrical aspects, as discussed in Anantharamaiah et al
(1993), are applied to restrict the acceptable combinations of \te,
\ne\- and \dhii. Finally, the models are used to compute the 
expected variation of line and continuum emission with frequency and
compared with observations. We show below that separate components
of ionized gas are required to explain the centimeter wavelength
(H92\a) and millimeter wavelength RRLs (H42\a, H40\a\- and H31\a).

\subsubsection{Models  based on the H92\a\- line and the Continuum}

Fig 8 shows three representative models that fit the observed H92\a\-
line.  The parameters of the three models are given in Table 7.  The
nature of the curves for other successful combinations of
\te, \ne\- and \dhii\- are similar to one of the three  curves shown in Fig 8, 
although the values of the derived parameters are different.  At
densities below about 500 \cmthree, the area covering factor of the
HII regions is unity. In other words, every line of sight through the
line emitting region intersects at least one HII region (\nhiilos $>1$
and $f_c$ =1 in Table 7), and thus the models at these densities are
similar to the uniform slab model discussed above. As seen in the
dashed curves in Fig 8, these low-density models are inconsistent
with the continuum spectrum below 1 GHz as well as the upper limit to
the RRL emission near 1.4 GHz.  Lower density models, which were
considered possible by Zhao et al (1996) based only on the H92\a\- and
higher frequency continuum data, are now ruled out.

At densities above about 5000 \cmthree, the area filling factor of the
HII regions is very small ($f_c <<$ 1) and thus they do not interfere
with the propagation of the non-thermal radiation. The continuum
spectrum will be non-thermal even at the lowest frequencies. The
dash-dot-dash curve in Fig 8 shows such a model (model C1) with the
parameters given in Table 7. In this model, although there are more
than $10^5$ HII regions, each $\sim$1 pc in diameter and optically
thick below $\sim$ 3 GHz, the total continuum emission continues to
have a non-thermal spectrum at lower frequencies. Thus these higher
density models cannot account for the observed turnover in the
continuum spectrum at $\nu <$ 500 MHz. These models are however
consistent with the upper limit to the RRL emission near 1.4 GHz.

A continuum spectrum, which is partially consistent with 
observations at low frequencies is obtained by models with densities
in the range $750 < n_e < 1500$ \cmthree.  An example is the solid
curve in Fig 8 (Model A1) with the parameters as given in Table 7. For
this model, the area covering factor $f_c =$ 0.7. In
other words 30\% of the non-thermal radiation propagates unhindered by
the HII regions and the remaining 70\% is subjected to the free-free
absorbing effects of the HII regions.  The net continuum spectrum thus
develops a break near the frequency at which $\tau_{HII} \sim 1$. At
much lower frequencies where 70\% of the non-thermal radiation is
completely free-free absorbed, the remaining 30\% propagates through with
its intrinsic non-thermal spectrum as shown in the solid curve in Fig
8. While this model accounts for the break in the spectrum near $\sim$
2 GHz, it fails to account for the complete turnover in the continuum
spectrum below 500 MHz. An additional thermal component with a
covering factor close to unity and a turnover frequency near $\sim$
300 MHz is required to account for the low-frequency spectrum. This
component is discussed in Section 4.2.3. Model A1 in Fig 8 is also
consistent with the upper limit to the RRL emission near 1.4 GHz.

In the models shown in Fig 8 and Table 7, between 25\% to 70\% of the
line emission arises due to external stimulated emission, i.e.
amplification of the background non-thermal radiation at the line
frequency due to non-LTE effects.  For lower densities, the fraction
of external stimulated emission is increased. For $n_e\sim 1000$
\cmthree\- (favored above), 70\% of the H92\a\- line is due to
stimulated emission. The other parameter of interest in Table 9 which
is related to the line emission mechanism is the non-LTE factor
$f_{nlte}$, the ratio of the intrinsic line emission from an
HII region if non-LTE effects are included to the line emission
expected under pure LTE conditions (i.e. $b_n = \beta_n =
1$). $f_{nlte}$ is in the range 1 to 3 for the models in Table
7. For lower  densities,  the non-LTE effect on the intrinsic
line emission from an HII region are reduced.  Thus, in the models with a
collection of HII regions, external and internal stimulated 
emissions vary with density in opposite ways. If the density is
increased, internal stimulated emission increases whereas external
stimulated emission decreases.

Two important parameters that can be derived from these  models
are the mass of the ionized gas \mhii and the number of Lyman
continuum (Lyc) photons \nlyc. The latter can be directly related to 
the formation rate of massive stars if direct absorption of Lyc photons
by dust is not significant.  Furthermore, predictions can be made for
the expected strengths of optical and IR recombination lines, which
can then be compared with observed values in order to estimate the
extinction. Some derived quantities are listed in Table 7. As seen in
this Table, although a range of \ne, \te, and \dhii\- values fit the
H92\a\- data, the derived value of \nlyc varies by less than a factor of
two. For the model labelled A1 in Fig 8, the total mass of ionized gas
\mhii = $3\times 10^7$ \msun\- and the number of Lyman continuum
photons \nlyc = $1.2 \times 10^{55}$ s$^{-1}$. 
For models that satisfy all the constraints, the derived
parameters \mhii\- and \nlyc\- increase if either \te\- is increased
or \dhii\- is decreased. On the other hand, as the density \ne\- is
increased, \mhii\- decreases, whereas \nlyc first increases and then
decreases. 

None of the models (at any density) that fit the H92\a\- line can
account for the observed H42\a, H40\a\- and H31\a\- lines. Although an
increase in line strength towards shorter wavelengths is predicted by
all the models (Fig 8), the expected integrated line flux density
falls short by almost an order of magnitude, well above any
uncertainty in the measurements. The model A1 in Fig 8 can explain
the H92\a\- line, the break in the continuum spectrum at 1.5 GHz and
is consistent with the upper limit to the H167\a\- line. Additional
components of ionized gas are required to explain the millimeter
wavelength RRLs and the turnover the continuum spectrum below 500
MHz.

\subsubsection{Models based on the H42\a\- line and the Continuum}

Since no model that fits the H92\a\- line could account for the
observed 3 mm and 1.2 mm lines, separate models, also based on a
collection of HII regions, were fitted to the H42\a\- line. Only
models with densities above $10^5$ \cmthree\- are consistent with the
mm wavelength RRLs. Three successful models, labelled A2, B2, and C2
are shown in Fig 9 and the parameters of the
models are listed in Table 8. As seen in Fig 9, the contribution of
this high density component to RRLs at centimeter wavelengths is
negligible since the HII regions are optically thick at these frequencies. This
high density component is thus detectable only in RRLs at millimeter
wavelengths.

In the models summarized in Table 8, the contribution to the line
emission from external stimulated emission is negligible ($<0.5\%)$.
On the other hand, enhancement of the line due to internal stimulated
emission within the HII regions is pronounced  and  sensitive to
the parameters of the HII regions (\ne, \te, and \dhii). The factor
$f_{nlte}$  range from 30 - 1300 for the models in Table 8. Because of the
sensitivity of the expected line strength to parameters of the HII
regions, the derived quantities (\mhii and \nlyc) can be well
constrained by the observed relative strengths of the H40\a\- and H31\a\-
lines.  Fig 9 indicates that the high density models can also be
distinguished by their contribution to the continuum flux density at
mm wavelengths.  

The Model labelled A2 in Fig 9 (and Table 8) gives a good fit to the
observed relative strengths of the H31\a, H40\a, and H42\a\-
lines. The solid line in Fig 9(b) shows the expected continuum
spectrum if only thermal emission from component A2 is added to the
non-thermal emission. Because of the low area filling factor ($f_c$)
of these components, there is no effect on the propagation of the
non-thermal radiation, although the HII regions are optically thick
below about 40 GHz.  Thermal emission from component A2 at 100 GHz is
only about 0.6 mJy.  In this model, the mass of ionized gas \mhii=
$3.6\times 10^3$ \msun\- which is negligible compared to the mass of
component A1, $6\times 10^7$ \msun\- (Table 7), required to explain
the H92\a\- line. The number of Lyman continuum photons \nlyc for the
high density component A2 is $3.5\times 10^{53}$ s$^{-1}$, about 3\%
of that for the lower density component A1 (Table 7).

The parameters of the high density component
that fit the mm wavelength RRLs are not unique. The alternative model
B2, shown as a dashed line in Fig 9, which is reasonably consistent
with the mm RRLs has a factor of ten higher \mhii\- and \nlyc\- (see
Table 8).  This model contributes a higher thermal continuum flux
density ($\sim 8$ mJy) at 100 GHz (see Fig 9). A better determination
of RRL and continuum flux densities are required to choose between
models A2 and B2.  As discussed in section 5.4, the high density
component provides important information about the star formation rate
during recent epochs in Arp 220.

\subsubsection{A Combined Model with Three Ionized Components}

In this section, we combine the best-fitting models presented in the
previous two sections and introduce a third component to account for
all the observations.  As discussed above, no single-density ionized
component is consistent with all the observations. The presence of
multiple components of ionized gas is to be expected since it is
extremely unlikely that a complex starburst region like Arp 220 could
consist of a single density ionized component. To construct a three
component model, we first select two models which provide good fits to
the H92\a\- and H42\a\- lines. We selected models which are labelled
A1 and A2 in Figs 8 and 9, respectively. The parameters of these
models are listed in Tables 7 and 8. The sum of the contributions from
these two models to the line emission can account for both H92\a\- and
H42\a\- and also is consistent with H40\a\- and H31\a\- line and the
upper limit to the H167\a\- line. Fig 10 illustrates the line and
continuum emission from these two components.  These two components
together can also account for the continuum spectrum above 1 GHz.  In
this model, the intrinsic non-thermal emission has a spectral index
$\alpha \sim -0.8$. Because of the presence of the thermal components,
the spectral index changes to $\sim -0.6$ in the range 2--20 GHz.  A
break in the spectrum is observed near 2 GHz which can be accounted
for by component A1.  This component, which has an area covering
factor of 0.7, becomes optically thick around 2 GHz and therefore
progressively absorbs about 70\% of the non-thermal radiation at lower
frequencies.  Component A2, which becomes optically thin above $\sim
40$ GHz, contributes very little to the continuum emission at any
wavelength. Thermal contribution to the continuum emission comes
mainly from component A1 and it becomes significant in comparison to
the non-thermal component above $\sim 20$ GHz. The thermal and
non-thermal emissions contributions are about equal at $\nu\sim$ 50
GHz. Since the thermal component exceeds the non-thermal component
above 50 GHz, the continuum spectrum becomes flatter at millimeter
wavelengths. In the continuum spectrum shown as a solid line in Fig
10b, thermal dust emission, which may have a significant contribution
even at 100 GHz, has not been included. Above 200 GHz, the total
continuum is dominated by dust emission (Rigopoulou et al 1996, Downes
and Solomon 1998). From the sub-millimeter continuum measurements by
Rigopolou et al (1996), we estimate that $\sim$20\% of the continuum
emission at 3 mm could be contributed by dust.

The observed continuum spectrum below a few hundred MHz is not
accounted for by the two thermal components A1 and A2 discussed above.
An additional free-free absorbing component with an emission measure
of a few times $10^5$ pc \cmsix\- and an area covering factor $f_c \sim
1$ is needed to account for the observed turnover. Furthermore, the
flux densities measured at 327 MHz and 150 MHz (Table 5) indicate that
the roll off in the spectrum is not steep enough to be accounted for by
a foreground thermal screen.  On the other hand, if the thermal gas is
mixed with the non-thermal emitting gas, then a shallower roll off is
expected.  It was possible to obtain a good fit to the observed
spectrum by adding a thermal component (mixed with the
non-thermal gas) with an emission measure EM = $1.3\times 10^5$,
\ne = 1000 \cmthree, \te = 7500 K and an area covering factor of unity.
The line and continuum emission of this component (shown as model D in
Fig 10) is very weak and is practically undetectable at any
frequency. The only observable aspect of this component is the
turnover in the continuum spectrum below a few hundred MHz.  The main
constraint for this component is its emission measure.  Densities
lower than about 500 \cmthree\-  can be ruled out as they are
inconsistent with the upper limit to RRLs near 1.4 GHz.  For the
model shown in Fig 10, \ne=1000 \cmthree\- and \te=7500 K which are same as
those of component A1.

All the parameters of this ionized component (model D) along with
those of components A1 and A2 and the sum of the three components,
where appropriate, are given in Table 9. Less than 10\% of the ionized
mass is in component D and $\sim$ 6\% of the Lyman continuum photons
arise from this component. Most of the mass (94\%) is in component
A1, accounting for $\sim 92$\% of the total Lyman continuum photons.
The mass in the high density component (A2) is negligible ($\sim
0.01$\%) and it accounts for about 3\% of the Lyman continuum photons.

Although the models given in Table 9 provide good fit to the observed
line and continuum data as seen in Fig 10, there are two aspects which
are not satisfactory: (1) Component A1, which contains the bulk of the
ionized gas at a density around 1000 \cmthree, is barely consistent
with the upper limit to the H167\a\- line. In fact, if this model is
correct, then the H167\a\- line should be detectable with a factor of
2-3 increase in sensitivity.  Although
an increase in density of this component will reduce the intensity of
the H167\a\- line (e.g. model C1 in Fig 8), the model will then fail
to account for the observed break in the continuum spectrum around 1.5
GHz.  (2) Model B2 in Fig 9, which is an alternative to A2 also
provides a reasonable fit to the observations, but has an order of
magnitude higher \mhii and \nlyc. This model slightly overestimates
the continuum flux density near 100 GHz but it provides  a good fit to the
mm wavelength RRLs.  As explained in Section 5.4, there are significant
implications to the star formation history of Arp 220 if
\nlyc in the high density component is as high as in component B2.
A resolution of the two difficulties mentioned here must await a firm
detection or determination of a more sensitive upper limit to the
H167\a\- line and a more accurate determination of the line and
continuum parameters at millimeter wavelengths. The validity or
otherwise of model B2 can be determined if several mm wavelength
RRLs and continuum are observed in the frequency range 100 to 300 GHz
and a proper separation of continuum emission by dust and ionized
gas is performed using the data.

\section{Discussion}

\subsection{Density}

As discussed in the previous section, only models with densities $>
\sim 10^3$ \cmthree\- can fit the recombination line data. Lower
density models are inconsistent with the upper limit to the
recombination line strength near 20 cm. While densities in the range
$ 10^3 - 2.5\times 10^4$ \cmthree\- fit the H92\a\- line, even
higher densities (\ne$\sim 1-5\times 10^5$ \cmthree) are required to
account for the millimeter wavelength recombination lines. These
results confirm the point made by Zhao et al (1996) that recombination
lines at different frequencies act as ``density filters''. Thus
multi-frequency RRL observations provides an excellent method of determining
the various density components in starburst regions.

That a substantial  fraction of the gas in the nuclear regions of
Arp 220 is at densities $>10^4$ \cmthree\- is evident from the high ${\rm
L_{HCN}/L_{CO}}$ ratio observed by Solomon, Downes and Radford
(1992). The total mass of H$_2$ gas at densities higher than $10^4$ \cmthree\-
is $>10^9$ \msun\- or more than 25\% of the dynamical mass of the two
nuclear components (Downes \& Solomon 1998). Less than $10^{-4}$ of this
gas needs to be in ionized form to account for the high density
ionized gas deduced from millimeter wavelength recombination lines.
The bulk of the ionized gas ($\sim 3\times 10^7$ \msun) which is
in component A1 (Table 9) is at a density $\sim 10^3$ \cmthree\- or
higher.

Electron densities inferred from [S {\small III}] and [Ne {\small
III}] IR line ratios observed with the ISO satellite typically yield lower
values ($n_e \sim 100 - 300$ \cmthree) (Lutz et al 1996).  We suggest
that this is a selection effect caused by the insensitivity of the [S
{\small III}] line ratios to densities outside the range $10^2 -
10^{3.5}$ \cmthree\- (Houck et al 1984). The upper limit to the
H167\a\- line near 1.4 GHz presented in Table 2, implies severe
constraints on the amount of low-density (i.e. \ne $<$ 500 \cmthree)
ionized gas present in Arp 220. Fig 11 shows the expected strength of
radio recombination lines as a function of frequency from ionized gas
with $n_e = 500$ \cmthree\- which is ionized by a Lyman continuum
luminosity equal to that deduced from Br\a\- and Br\g\- recombination lines observed
with ISO (i.e. \nlyc = $1.3 \times 10^{55}$ s$^{-1}$, G\"{e}nzel et al
1998). The predicted strength of the H167\a\- recombination line is
five times the 3$\sigma$ upper limit in Table 2. Thus the only way to
consistently explain the ISO and radio results is if the IR
recombination lines (Br\a\- and Br\g) also arise in higher density gas
(i.e. $n_e > 10^3$ \cmthree).  The three component model in Table 9
shows that the total number of Lyman continuum photons, \nlyc, 
is $\sim 1.3\times 10^{55}$ photons s$^{-1}$.  This value
of \nlyc\- is equal to the intrinsic Lyman continuum luminosity
inferred from IR recombination lines observed with the ISO satellite
(Sturm et al 1996, Lutz et al 1996, G\"{e}nzel et al 1998). The simplest
reason for this correspondence is that almost all of the IR
recombination lines also arise in the same gas with density \ne $ > 10^3$
\cmthree.  That still leaves the question why [S III] line ratios
observed with ISO indicates a density of only a few hundred \cmthree,
although they are sensitive to up to ten times higher densities. It is
possible that if dust extinction is very high (see next section),
[SIII] lines are not reliable indicators of density.

\subsection{Extinction}

The predicted flux of the Br\a\- line from the three component model in
Table 9 is $1.0\times 10^{-15}$ W~m$^{-2}$. Sturm et al (1996)
report a measured flux of $2.1 \times 10^{-16}$ W~m$^{-2}$ from ISO
data. Thus extinction of Br\a\- is $A_{Br\alpha} \sim 1.7$ which
correspond  to a $V$ band extinction $A_V \sim 45$. Similar calculation
based on the predicted Br\g flux of $3.2 \times 10^{-16}$ W~m$^{-2}$
and a measured flux of $5.9\times 10^{-18}$ W~m$^{-2}$ (Goldader et al
1995) yield $A_{Br\gamma} \sim 4.3$ and $A_V \sim 41$. These derived
values of $A_V$ are comparable to the values obtained using various IR
line ratios by Sturm et al (1996). This similarity in the derived
$A_V$ values provides further evidence that the IR recombination lines
of hydrogen observed by ISO arise in high density gas. The radio
recombination line data confirms the high extinction values found
from ISO spectroscopy (Sturm et al 1996, G\"{e}nzel et al 1998) and provides
support to the idea that Arp 220 is powered by a massive starburst (see
below).

\subsection{\nlyc and Star Formation Rate (SFR)}

One of the derived quantities of the three component model in Table 9
is the production rate of Lyman continuum photons \nlyc. Lyman
continuum (Lyc) photons are produced by massive O and B stars during their
main sequence phase.  Since these stars have a relatively short main
sequence life time of $< 10^7$ years, the derived value of \nlyc can
be related to the formation rate of O and B stars if direct absorption
of Lyc photons by dust is not significant. If an initial mass
function (IMF) is adopted, then the formation rate of OB stars can be
used to derive the total star formation rate (SFR) (e.g. Mezger
1985). The derived SFR is both a function of the adopted IMF and
assumed upper and lower  mass limits of stars that are formed 
($m_u$ and $m_l$). It is thought that in the case of induced star
formation, triggered for example by an external event such as a
galaxy-galaxy interaction or a merger, the lower mass cut off in the
IMF may be a few solar masses (Mezger 1987). In spontaneous
star formation in a quiescent molecular cloud, the lower mass cut off
may be $<0.1$ \msun\- set by theoretical considerations (Silk 1978). In
the following discussion we use $m_l$ = 1 \msun\- and $m_u$ = 100 \msun,
the latter corresponding to an O3.5 star. Using these mass limits and
assuming a constant rate of star formation (over the life time of O
stars), we get from the formulae given by Mezger (1985)
\begin{equation}
N_{Lyc} = 5.4 \times 10^{52} \times {\bf \Psi}_{OB}~~ {\rm s^{-1}},
\end{equation}
where ${\bf \Psi}_{OB}$ (\msun\- yr$^{-1}$) is the SFR averaged over
the lifetime of the OB stars. In eqn 1, the IMF proposed by Miller
and Scalo (1978) has been used. Using a single power-law IMF
of Salpeter (1955) results in a factor of $\sim 3$ increase
in \nlyc. A reduction in the lower mass limit $m_l$ to 0.1 \msun\-
decreases \nlyc by $\sim 2$. The value of \nlyc is less sensitive
to the upper mass limit in the range 50-100 \msun. 

The sum of \nlyc for the three components in Table 9 is $1.3\times
10^{55}$ s$^{-1}$. An examination of the various models discussed in
the previous section indicate that the uncertainty in this value is
likely to be less than a factor of two. This value of \nlyc is
consistent with the results from ISO based on NIR recombination lines
(Sturm et al 1996 and G\"{e}nzel et al 1998) and also with the lower limit
of $1 \times 10^{55}$ s$^{-1}$ deduced by Downes and Solomon (1998)
based on their estimated thermal continuum at 113 GHz.

The derived SFR based on equation (1) and the total \nlyc in Table 9
is $\sim$ 240 \msun\- yr$^{-1}$. The SFR in Arp 220 is thus about two
orders of magnitude higher than in the Galaxy and may be the
highest SFR derived in any normal, starburst or ultra-luminous galaxy
(e.g. Kennicutt et al 1998). If the Salpeter-IMF is used and the upper mass
limit is reduced to 60 \msun, then the SFR is reduced to 90 \msun\- yr$^{-1}$.

The high SFR in Arp 220 is likely a consequence of the large gas
content in the nuclear region with high volume and surface
densities. Downes and Solomon (1998) have estimated that within a
radius of about 500 pc, which includes the east and west nuclei as
well as a portion of a gas disk that surrounds them, the total mass of
H$_2$ gas is $\sim 4 \times 10^9$ \msun. For the same region, Scoville
et al (1997) obtain a higher mass of $\sim 8 \times 10^9$ \msun . For
the discussion here, we take a mean of the two, $6\times 10^9$
\msun. The average volume density of the H$_2$ is $\sim 250$
\cmthree\- and the average gas surface density $\mu_g \sim$ 6000
\msun\- pc$^{-2}$. A detailed study of star formation rates in normal
and ultra-luminous galaxies by Kennicutt (1998) has shown that SFR per
unit area can be fitted to a Schmidt law of the form $\Sigma_{SFR} =
2.5\times 10^{-4} \mu_g^{1.4}$ \msun\- yr$^{-1}$ kpc$^{-2}$. This
relation yields an SFR of 38 \msun\- yr$^{-1}$ for $\mu_g$ = 6000
\msun\- pc$^{-2}$, which is a factor of six lower than deduced
above. In his compilation of starburst properties of luminous
galaxies, Kennicutt (1998) uses $\mu_g$ = 5.8$\times 10^4$ \msun\-
pc$^{-2}$ for Arp 220 and an SFR of 955 \msun\- yr$^{-1}$ kpc$^{-2}$.
This high value of $\mu_g$ is the peak surface density in the inner
most region obtained by Scoville et al (1997) and it is about a factor
of 10 higher than the average surface density over a $\sim$kpc$^2$
area.  Thus, when averaged over a kpc$^2$ region, the SFR predicted by
Schmidt law with an exponent of 1.4 falls short of the value deduced
above (240 \msun\- yr$^{-1}$) by about a factor of six.  Although
other parameters such as the upper and lower mass limits and the shape
of the IMF could be adjusted to lower the deduced SFR, it is unlikely
to reduce it to the value predicted by the empirical Schmidt law
obtained by Kennicutt (1998).

In a study of star forming regions in normal galaxies, Viallefond et al
(1982) found a relation between $M_{gas}$, \nlyc and density $n$ of the form
\nlyc = $7.9\times 10^{44}~M_{gas}~n^{0.36}$. Using 
$M_{gas} = 6\times 10^9$ \msun\- and $ n= 250$ \cmthree\-  (the
average molecular density), this relation predicts \nlyc = $3.3\times
10^{55}$ s$^{-1}$. Although this value is about a factor of 2.5 higher
than the derived total \nlyc in Table 9, these values are comparable
given the uncertainties. This result shows that the starburst in
Arp 220 behaves  as expected from scaling the known properties of
star forming regions in normal galaxies.


\subsection{Star Formation at Recent Epochs}

As shown in Section 4, the observed intensities of the
millimeter wavelength RRLs H31\a, H40\a\- and H42\a\- in Arp 220 can only be
explained by ionized gas at a high density of $\sim 2.5\times 10^5$
\cmthree. The presence of these high density HII regions which account
for about 3\% of the total \nlyc of $1.3\times 10^{55}$ s$^{-1}$, can
be used to derive the  star formation rates at recent epochs. Since the
high density compact HII regions are relatively short-lived ($\tau_{HII} \le
10^5$ yrs), the \nlyc value corresponding to these regions will indicate
the SFR averaged over the life time of  the HII regions rather than the
main-sequence life time of O, B stars which ionize them. Based on
equation (1), and approximating the Lyc photon production rate of
OB stars to be constant during their life time, we can write
\begin{equation}
{\bf \Psi}_{HII} = {\bf \Psi}_{OB} \times \left[\frac{\tau_{OB}}{\tau_{HII}}\right]
                   \times \left[\frac{N_{Lyc,HII}}{N_{Lyc,OB}}\right].
\end{equation}
The compact, high density phase of an HII region is short lived because
the HII region expands as it is over-pressured with respect to the
surroundings. The expansion proceeds at the sound speed $c_i$ ($\sim 10$ \kms)
and approximately follows the relation
\begin{equation}
r(t) = r_i\left[1+\frac{7c_it}{4r_i}\right]^{4/7}
\end{equation}
Spitzer (1968), where $r_i$ is the initial size of the HII regions and
$r(t)$ is its size after a time $t$. The HII regions of component A2 in Table
9 have a density of $2.5\times 10^5$ \cmthree. The density of these HII
regions will drop to about 1000 \cmthree\- and become indistinguishable from
component A1, if they expand to $\sim 6$ times their initial size. Taking
the initial size to be 0.1 pc, as given in Table 9, equation (3) gives
$t =10^5$ yrs. If the initial size is smaller, then the time scale is also
shorter. (For a single O5 star, the initial size of the Stromgren sphere
is $\sim$0.05 pc if the density is  $2.5\times 10^5$ \cmthree.) Using
$\tau_{HII} = 10^5$ yrs, $\tau_{OB} = 3\times 10^6$ yrs, ${\bf\Psi}_{OB} $
= 240 \msun\- yr$^{-1}$ and using \nlyc values from Table 9, equation 2 gives
the SFR averaged over the life time of HII regions  ${\bf \Psi}_{HII} =$ 194
\msun\- yr$^{-1}$. This rate is similar to the SFR averaged over $\tau_{OB}$.
This result indicates that the starburst in Arp 220 is an on-going process
with a minimum age exceeding $\tau_{OB}$. Other evidence (see below) indicates
that the age of the starburst is probably much longer.

The above calculation demonstrates that a reliable measurement of the
number of Lyman continuum photons in high density HII regions can lead
to a determination of the ``instantaneous'' SFR. Such a measurement
seems possible through RRLs and continuum at millimeter wavelengths
which are sensitive to the dense component.  The actual value of the
instantaneous SFR determined above (194 \msun\- yr$^{-1}$) can be 
almost a factor of ten  higher or about a factor of two 
 lower  since the value of ${N_{Lyc,HII}}$ in equation 2 is
uncertain by those factors (see Table 8). If the ten times higher value of
$N_{Lyc,HII} \sim 4.6\times 10^{54}$ s$^{-1}$ corresponding to model B2 in
Table 8 is established through further measurements, then the implied
SFR at recent epochs will be ten times the average value. Such a 
result will fundamentally alter our picture of the starburst process
in Arp 220. Instead of being bursts of constant but high SFR
(approximately 10-100 times the Galactic rate) over $10^6 - 10^7$ yrs,
starburst in Arp 220 will be multiple events of short duration ($10^5 - 10^6$
yrs) but at extremely high SFR (100-1000 times the Galactic rate).
Further continuum and RRL measurements at millimeter wavelengths are
required to clarify this aspect.

\subsection{\nlyc and IR-Excess}

In dense galactic HII regions, the ratio $\frac{L_{IR}}{L_\alpha}$ is
in the range 5-20 which is known as the IR-excess (IRE) (Mezger et al
1974, Panagia 1978, Mathis 1986). $L_{IR}$ is the observed infrared
luminosity. $L_\alpha$ is the luminosity of Lyman-$\alpha$ photons
deduced from observations of the ionized gas using recombination lines
or radio continuum and assuming that all the Lyman continuum photons
which ionize the gas are eventually converted to L\a\- photons
(i.e. $N_{L_\alpha}$=\nlyc).  These L\a\- photons are then absorbed by
dust. If this is the only source of heating the dust in an ionization
bounded HII region, and since dust re-radiates most of the absorbed
energy in the IR, then the expected IRE is $\sim$1. If IRE $>1$, then
the dominant mechanism for heating the dust in HII regions contains 
a major contribution other
than  $L\alpha$ heating. Two possible sources of heating are:
(1) direct absorption of Lyc photons by dust and (2) absorption of
photons long ward of the Lyman limit. Lyc photons which are directly
absorbed by dust are not counted in the \nlyc derived from RRL or
continuum observation of the ionized gas.  Therefore, if direct
absorption of Lyc photons is the main contributor to the IRE, then the
SFR derived using eqn 1 will be an underestimate by a factor
$\sim$IRE. On the other hand, heating of dust by photons long ward of
the Lyman limit (which are produced mainly by lower mass, non-ionizing
stars) can account for the IRE without a corresponding increase in the
SFR. Viallefond  (1987) has shown that a population of non-ionizing
stars formed continuously or in a burst with a Miller-Scalo IMF can
indeed explain an IRE of $\geq$10. In Arp 220, $L_{IR} = 
1.3 \times 10^{12} L_\odot$ and $L_{L_\alpha}
= 5.5\times 10^{10}L_\odot$ using $N_{L_\alpha}$ = \nlyc =
$1.3\times 10^{55}$ s$^{-1}$. Thus for Arp 220, IRE $\sim$ 24. Based
on this value of IRE we make a few deductions about the star formation
history in Arp 220.

\subsubsection{Starburst or AGN?}

In giant and super-giant HII regions of the normal galaxy M33, the
value of IRE is in the range 3 to 8 and the global value of IRE is
$\sim 14$ (Rice et al 1990). In the Galaxy, IRE is typically $\sim$ 7
in compact HII regions and $\sim 5$ in extended HII regions (Mezger
1987). G\"{e}nzel et al (1998) have computed a modified form of IRE (they
use $L_{Bol}/L_{Lyc}$) in starburst galaxies observed by ISO.
Converting their ratios to the above definition of IRE
(i.e.$\frac{L_{IR}}{L_\alpha}$), the mean value of IRE in starburst
galaxies is $\sim 25$ with values ranging from 12 to 45.  Similar
computation made for AGN dominated sources show that IRE$ \sim 45-65$
(G\"{e}nzel et al 1998). Thus IRE in Arp 220 ($\sim 24$) is very similar
to the values observed in starburst galaxies and slightly higher than
in star forming regions in the Galaxy and in M33. IRE in Arp~220 is
significantly lower than in AGN dominated sources.  The high infrared
luminosity of Arp 220 can therefore be entirely accounted for by
processes that operate in starburst regions. We therefore conclude
that Arp 220 is powered entirely by a starburst. There is no
significant contribution from an ``active'' nucleus to the observed
high IR luminosity in Arp 220.

\subsubsection{IMF, Age of Starburst and Star Formation Efficiency in Arp 220}

The fact that the IRE for Arp 220 is not significantly different from the IRE
in star forming regions in the Galaxy and in M33 also implies that the
starburst in Arp 220 has an IMF which is not unusual. If, for example,
IRE$\sim$1, then there is no contribution to heating of dust by
non-ionizing photons which would imply that the burst is mainly
producing massive stars and thus $m_l > \sim$ 5 \msun. This is not the
case in Arp 220. On the other hand if IRE$>>$10, then it would imply
one of the following: (1) the IMF is truncated at the upper end ($m_u
< 10-20$ \msun\- or so) which makes the heating of dust by
non-ionizing photons even more pronounced, leading to a much larger
IRE or (2) the starburst ended about $5\times 10^6$ years ago which
would reduce \nlyc in relation to non-ionizing photons, again
resulting in a larger IRE or (3) an AGN is contributing predominantly
to the heating and IR emission. These three possibilities are also
ruled out in Arp 220. 

Another implication of IRE $\sim$ 24 in Arp 220, which is
significantly higher than in young star forming regions, is that the
starburst in Arp 220 must be much longer in duration than the lifetime
of OB stars (i.e. $t_{SB} > 10^7$ yrs). As mentioned above, in
younger star forming regions such as the super-giant HII region NGC
604 in M33 and in compact HII regions in the Galaxy, the IRE is
significantly smaller ($\sim 5-7$, Mezger 1985, Rice et al 1990).  To
illustrate that a relatively high value of IRE implies a longer
duration of starburst, we consider a simple model of continuous star
formation discussed in Viallefond and Thuan (1983) and the tabulation
of the derived parameters in Viallefond (1987). In this model, for
an assumed IMF with lower ($m_l$) and upper ($m_u$) stellar mass limits, 
\nlyc and \lbol\- are computed as a function of duration of star formation
($\tau_{csf}$) for a SFR of 1 \msun\- yr$^{-1}$.  These values, taken
from Viallefond (1987), are tabulated in Table 10 for a Miller-Scalo
IMF with $m_u = 100$ \msun\- and $m_l = 1$ \msun. The results would be
similar for $m_l = 0.1$  \msun. In this model, a steady state
production rate of Lyc photons is reached after $\sim 5\times 10^6$
yrs. The steady state value of
\nlyc is $3.7 \times 10^{52}$ ph s$^{-1}$/(\msun\- yr$^{-1}$) and
implies a SFR of $\sim$ 350 \msun\- yr$^{-1}$ since the derived value
of \nlyc = $1.3\times 10^{55}$ ph s$^{-1}$. This SFR is consistent
with the value derived in Section 5.3. Table 10 also shows that in
this model, the observed IRE$\sim$ 24 in Arp 220 is reached at
$\tau_{csf} \sim 1.5\times 10^7$ yrs.

Finally, a continuous SFR of 350 \msun\- yr$^{-1}$ for a duration of
$1.5\times 10^7$ yrs converts $\sim 5\times 10^9$ \msun\- of gas into
stars. Thus the mass of stars formed is approximately equal to the
present mass of gas ($\sim 6\times 10^9$ \msun\-).  In a closed-box
model (i.e. with no fresh gas added to the system from outside), the
star formation efficiency (SFE) in Arp 220 is 50\%, much higher than
the average SFE of a few \% in Galactic star forming regions
(Elmegreen 1983, Myers et al 1986). The high SFE ($\sim 50$\%) and the
high SFR ($\sim$300\msun\- yr$^{-1}$) in Arp 220 and its corresponding
high luminosity ($L_{IR} = 1.3 \times 10^{12}$ \lsun) must be a result
of confinement of a large quantity of gas (M$_{gas} \sim 6\times 10^9$
\msun) in a relatively small volume ($\sim$kpc$^3$) with a high
average density of 250 \cmthree. At the East and West peaks of Arp
220, about $10^9$ \msun\- of gas (i.e. about the mass of gas in the
Galaxy) is confined to a region $\sim$100 pc in size leading to a much
higher {\em average} density of  $\sim 10^4$
\cmthree. The high concentrations of large quantities of gas in Arp
220 have given rise to a starburst with an efficiency and SFR that is
consistent with known properties of star forming regions in normal and
starburst galaxies.

\section{Summary}

We have presented observations of radio recombination lines and
continuum emission in Arp 220 at centimeter and millimeter
wavelengths. We have showed that to explain both the observed variation of
recombination line intensity with quantum number and also the observed
continuum spectrum in the frequency range 0.15 - 113 GHz, three
components of ionized gas with different densities and covering factors
are required. 

The bulk of the ionized gas is in a component (A1) with an electron density
$\sim 1000$ \cmthree. The total mass in this component is $3\times
10^7$ \msun. This component of ionized gas consists of $\sim 2\times
10^4$ HII regions each $\sim$5 pc in diameter.  This ionized component
produces detectable recombination lines at centimeter wavelengths and
substantially modifies the non-thermal continuum spectrum at these
wavelengths.  The ionized gas in this component becomes optically
thick below $\sim 1.5$ GHz and partially absorbs the non-thermal
radiation since its area covering factor is $\sim 0.7$. The intrinsic
spectral index of the non-thermal radiation in Arp 220 is $\alpha \sim
-0.8$ which is modified to an observed value of $\sim -0.6$ in the
5-10 GHz range due to the presence of this ionized gas. At $\nu$ = 5 GHz,
the total observed flux density of Arp 220 is $\sim 210$ mJy, of which
175 mJy is non-thermal and the remaining $\sim 35$ mJy is thermal
emission from component A1. The other two thermal components (see
below) produce negligible continuum emission at this frequency.

A second component (labelled A2) at a density of $\sim 2.5\times 10^5$
\cmthree\- with a mass of only $3.6\times 10^3$ \msun\- accounts for the
recombination lines observed at millimeter wavelengths and it is not
detected in RRLs at centimeter wavelengths. Component A2 consists of
$\sim 10^3$ HII regions, each 0.1 pc in diameter; these regions are
optically thick below 40 GHz. The area covering factor of
component A2 is small ($\sim 10^{-5}$) and thus it does not affect the
intrinsic spectrum of the non-thermal radiation at centimeter
wavelengths. An alternative model which also fits the data has an
order of magnitude higher mass at a similar density. Improved
measurements of line and continuum parameters at millimeter wavelengths
are required to choose  the correct model.

A third component (labelled D) of ionized gas with an emission measure
of 1.3$\times 10^5$ pc \cmsix\-, density $>500$ \cmthree, mixed
with the non-thermal gas, is needed to account for the observed
turnover in the continuum spectrum below 500 MHz. The mass in this
ionized component is $2\times 10^6$ \msun\- and requires about
$8\times 10^{53}$ Lyc photons s$^{-1}$.

The total mass of ionized gas in the three components is $\sim
3.2\times 10^7$ \msun\- requiring $3\times 10^5$ O5 stars to
maintain the ionization. The total production rate of Lyman continuum
photons, \nlyc = $1.3\times 10^{55}$ s$^{-1}$ of which $\sim$92\% is
used by component A1, $\sim$3\% by component A2 and $\sim 5$\% by
component D. \nlyc deduced from radio recombination lines is
consistent with the values obtained using NIR recombination lines
observed with ISO.  A comparison of the predicted strengths of Br\a\-
and Br$\gamma$ lines summed over all the three components with
observed values shows that the V band extinction due to dust, $A_V
\sim 45$ magnitudes. This value of $A_V$ is consistent with the
results from ISO observations.

On the assumption of continuous star formation with an IMF proposed by
Miller and Scalo (1978) and assuming a stellar mass range of 1-100
\msun,  the deduced value of \nlyc implies a SFR of 240 \msun\- yr$^{-1}$
averaged over the mean main sequence life time of O, B stars ($\sim
3\times 10^6$ yrs). The dense HII regions of component A2 are short
lived ($\tau_{HII} \sim 10^5$ yrs) since they are highly
over-pressured. Thus the value of \nlyc corresponding to this
component is related to very recent rate of star formation,
i.e. averaged over a time scale $\tau_{HII} \sim 10^5$ yrs. The
deduced recent SFR is similar to the average over $\sim 10^7$ yrs; thus
the starburst in Arp 220 is an on-going process. An alternative model,
which cannot be excluded on the basis of the available data, predicts an
order of magnitude higher mass in the dense HII regions and
correspondingly high SFR at recent epochs (i.e. $t<10^5$ yrs). If this
alternative model is confirmed through further observations at
millimeter wavelengths, then the starburst in Arp 220 consists of
multiple episodes of very high SFR (several thousand \msun\-
yr$^{-1}$) of short durations ($\sim 10^5$ yrs).

Finally, based on the value of \nlyc deduced from RRL and continuum
data, the IR-excess (i.e. the ratio $L_{IR}/L_{L_\alpha}$) in Arp 220 is
$\sim 24$, comparable to the values found in starburst galaxies. This
similarity in IR-excess implies that Arp 220 is most likely powered
entirely by a starburst rather than an AGN.  A comparison of the IRE in
Arp 220 with the IRE in star forming regions in the Galaxy and in M33
indicates that the starburst in Arp 220 has a normal IMF and a duration
much longer than $10^7$ yrs. If no in-fall of gas has taken place
during this period, then the star formation efficiently (SFE) in Arp
220 is $\sim 50$\%. The high SFR and SFE in Arp 220 is a consequence
of concentration of a large mass in a relatively small volume and is
consistent with known dependence of star forming rates on mass and
density of gas deduced from observations of star forming regions in
normal galaxies.

\acknowledgments  

The National Radio Astronomy Observatory is a facility of the National
Science Foundation operated under cooperative agreement by Associated
Universities, Inc.

\clearpage

\begin{table}
{\bf Table 1} Observing Log  - VLA \\

\begin{tabular}{lll}
\hline
Parameter                                          &   H92\a\- Line              &  H167\a\- and H165\a\- Lines \\
\hline
Date    of  Observation                    & 1 Aug 98                    & 25 Mar 98 \\
Right Ascension (B1950)               & 15 32 46.94                & 15 32 46.94\\
Declination (B1950)                       & 23 40 08.4                  & 23 40 08.4 \\
VLA Configuration                           & B                                   & A                \\
Observing duration (hrs)                & 12                                 &   7 \\
Range of Baselines (km)                & 2.1 - 11.4                  & 0.7 - 36.4 \\
Observing Frequency(MHz)         & 8164.9                          & 1374.0 \& 1424.4\\
Beam  (Natural weighting)          & $1.1''\times 0.9''$     & $1.8'' \times 1.6''$ \\
Bandwidth (MHz)                            & 50                                & 6.25\\
No. of Spectral Channels        & 16                                 & 32\\
Number of IFs                                   & 1                                   & 4\\
Center $V_{Hel}$ (\kms)              & 5283                             & 5555\\
Velocity coverage   (\kms)             & 1360                              & 1220\\
Velocity resolution   (\kms)            & 230                               & 84\\
Amplitude Calibrator                       & 3C 286                         & 3C 286\\
Phase Calibrator                                & 1600+335                   & 1511+238 \& 1641+399\\
Bandpass calibrator                          & 1600+335                  & 3C 286\\
RMS noise per Channel ($\mu$Jy) & 40                             & 130\\
\hline
\end{tabular}
\end{table}


\begin{table}
{\bf Table 2} Observed  Continuum and Line Parameters -  VLA \\

\begin{tabular}{lll}
\hline
Parameters                                           &  H92\a\- line                      & H167\a\- + H165\a\-  lines \\
\hline 

Peak Line flux density (mJy)          & 0.6 $\pm$ 0.1                & $<$ 0.25 (3$\sigma$) \\
Continuum Flux over\\  \hspace{2pt}  
 the Line region  (mJy)                     & 132$\pm$ 2            & -- \\
Total Continuum Flux (mJy)              & 149$\pm$ 2          & 312 $\pm$ 3\\
Angular size of line region               &  $2.5'' \times 1.5''$     & -- \\
Central V$_{Hel}$   (\kms)            & 5450$\pm$20                                 & --\\
Line width (FWHM, \kms)              & 363 $\pm$45                & 360 (assumed) \\
Integrated Line flux ( W m$^{-2}$) & 8.0 $\pm$1.5$\times 10^{-23}$ & $< 0.5 \times 10^{-23}$\\
\hline
\end{tabular}
\end{table}

\begin{table}
{\bf Table 3} Observing Log - IRAM 30 m\\

\begin{tabular}{rcccccccc}
\hline
Observing & Aperture  &$S_{\nu ,{\it b}}/{T_A}^{*}$ & Beam       & \multicolumn{2}{c}{Bandwidth} & \multicolumn{2}{c}{Channel}  &Integ.\\
Frequency&efficiency  &                                                          &(FWHM)   &       &                                                    &\multicolumn{2}{c}{separation} & time \\
GHz           &                     &      Jy/K                                           & arcsec       &                MHz   &  \kms   &         MHz   & \kms          &Hours\\
\hline 
206.669  & 0.39     & 8.6                       &  11.6           & 512   & 742                 & 1.00  & 1.45                 &2.8  \\
 97.220  & 0.58     & 6.1                       &  24.7           & 510   & 1576                & 1.25  & 3.85                 &2.8  \\
 84.128  & 0.62     & 5.7                       &  28.5           & 510   & 1821                & 1.25  & 4.45                 &2.8  \\
\hline
\end{tabular}
\end{table}

\begin{table}
{\bf Table 4} Observed Line and Continuum Parameters -  IRAM 30 m\\

\begin{tabular}{lrrrccc}
\hline
 Line      & Observed  &Line flux              & V$_{Hel}$    & Velocity           & Relative  & S$_{Contm}$ \\
               & frequency  &                              &                           & dispersion        & flux      \\
\hline
              & GHz            & Jy \kms & \kms &  \kms &  Jy/Jy    &       Jy  \\
\hline
H31$\alpha$& 206.67& $ > 24.0 \pm 1.4$ &        &                                      & 1.00 &  $121\pm 17$  \\
H40$\alpha$&  97.22 & $   12.2 \pm 1.6$ & $5513$ &  $179$                  & 0.28    & $61 \pm 10$\\
H42$\alpha$&  84.13 & $    7.9 \pm 0.9$ & $5424$ &  $210$                   & 0.24   & $\le 72      $ \\
\hline
\end{tabular}
\end{table}

\begin{table}
{\bf Table 5} Continuum Flux Densities of Arp 220\\

\begin{tabular}{rrc}
\hline
$\nu$ (GHz)  &  S$_{\nu}$  (mJy)  & Reference \\
\hline

0.15            &  260$\pm$ 20             & 1 \\
0.325          &  380$\pm$ 15             &  2\\
1.40            &  312$\pm $03             & 2    \\   
1.40            &  295$\pm$05              & 3      \\
1.60            &  332$\pm$04              & 3 \\
2.38            &  312$\pm$16              & 1 \\
4.70            &  210$\pm$02              & 3 \\
8.16            &  149$\pm$02              & 2 \\
10.7            & 130$\pm$15               & 1\\
15.0            & 104$\pm$02               & 3\\
22.5            &   90$\pm$06               &3 \\
97.2            &   61$\pm$10               & 2 \\
113.0          &   41$\pm$08               & 4\\
206.7          & 121$\pm$17               & 2\\
226.4           & 175$\pm$35              & 4\\
\hline
\end{tabular}
\tablerefs{(1) Sopp \& Alexander (1991), (2) This paper, (3) Zhao et al (1996)
(4) Downes \& Solomon (1998)}

\end{table}

\begin{table}
{\bf Table 6} Parameters of  Uniform Slab Models \\

\begin {tabular}{lccc}
\hline
Parameter &  Model A & Model B & Model C  \\
\hline

  T$_e$ (k)                     &   7500  &7500  & 7500  \\
  N$_e$ (\cmthree)          &   10 &      100 &     1000  \\   
  Path length (pc)     &  2.5$\times 10^4$&   120     &  2.1         \\
  EM (pc \cmsix)     &    2.5$\times 10^6$ & 1.2$\times 10^6$  & $2.1\times 10^6$ \\
  S$_{thermal}$ (5 GHz) (mJy)&  22.1     &  11.1     &  19.0  \\
  S$_{non-th}$ (5 GHz)   (mJy)     & 190.7 &      196.     &  192. \\
  $\alpha_{non-th}$            & --0.56    & --0.580   &  --0.57   \\
  Stimulated Emission (\%)      &   75.0   &    86.7  &     73.3   \\
  $\tau_c$ (5GHz)        &  0.04  & 0.02 &  0.034 \\
  \mhii (\msun)        &   4.0 $\times 10^ 9$ &  2.0$\times 10^8$ & 3.5$\times 10^7$\\
  N$_{Lyc }$(s$^{-1}$)         &  1.6$\times 10^{55}$  & 7.9$\times 10^{54}$   &1.4$\times 10^{55}$\\
  N$_{O5}$                 &  3.4$\times 10^ 5$  & 1.7$\times 10^5$ &  2.9$\times 10^5$   \\
  F$_{Br\alpha}$ (W m$^{-2}$)& 1.2$\times 10^{-15}$ & 6.0$\times 10^{-16}$ & 1.1$\times 10^{-15 }$\\
  F$_{Br\gamma}$ (W m$^{-2}$) &3.9$\times 10^{-16}$  &1.9$\times 10^{-16}$ &  3.3$\times 10^{-16}$ \\
\hline
\end{tabular}
\end{table}

\begin{table}
{\bf Table 7} Parameters of  HII-regions  Model  Based on the H92\a\- Line\\

\begin {tabular}{lccc}
\hline
Parameter &  Model A1 & Model B1 & Model C1  \\
\hline

  T$_e$ (K)                 & 7500               & 7500               &  7500    \\
  N$_e$ (\cmthree)          & 1000               & 500                & 5000 \\
  size (pc)                 & 5.0                & 2.5                & 1.0 \\
  N$_{HII}$                 & 1.9$\times 10^4$   & 5.9$\times 10^5$   & 1.2$\times 10^5$  \\
  N$_{HIIlos}$              & 0.7                & 5.6                & 0.2    \\
  Area covering factor $f_c$& 0.7                & 1.0                & 0.2 \\
  Volume Filling factor $f_v$ & 4.3$\times 10^{-3}$& 1.7$\times 10^{-2}$& 2.2$\times 10^{-4}$\\
  $\tau_c$(8.3GHz)          & 0.03               & 0.003              & 0.14    \\
  $\tau_L$(8.3GHz)          & --0.13             & --0.02             & --0.25  \\
  $b_n$                     & 0.947              & 0.923              & 0.980        \\
  $\beta_n$                 &--45.5              & --63.2             & --17.9       \\
  Stimulated emission (\%)  & 54                 &  71                &  24.8     \\
  $f_{nlte}$                & 1.7                &  1.0               &  2.6       \\
  S$_{thermal}$ (5GHz)(mJy) & 31                 &  31                &  33       \\
  S$_{non-th}$  (5GHz)(mJy) & 176                &  176               &  175      \\
  $\alpha_{non-th}$         &--0.76              & --0.76             & --0.84       \\
 \mhii  (\msun)             &3.0$\times 10^7$    &5.9$\times 10^7$    & 7.5$\times 10^6$   \\
  N$_{Lyc}$  (s$^{-1}$)     &1.2$\times 10^{55}$ &1.2$\times 10^{55}$ & 1.5$\times 10^{55}$ \\
  N$_{O5}$                  &2.5$\times 10^5$    &2.5$\times 10^5$    & 3.1$\times 10^5$    \\
  F$_{Br\alpha}$(W m$^{-2}$)&9.2$\times 10^{-16}$&9.1$\times 10^{-16}$& 1.1$\times 10^{-15}$ \\
  F$_{Br\gamma}$(W m$^{-2}$)&2.9$\times 10^{-16}$&2.9$\times 10^{-16}$& 3.6$\times 10^{-16}$ \\
  N$_{O5}$ per HII region   &     0.42           &      13.6          &      2.7  \\    
\hline
\end{tabular}
\end{table}

\begin{table}
{\bf Table 8} Parameters of  HII-regions  Model  Based on the H42\a\- Line\\

\begin {tabular}{lccc}
\hline
Parameter &  Model A2 & Model B2& Model C2  \\
\hline

  T$_e$ (K)                      &  7500               &  $10^4$             &  $ 10^4$   \\
  N$_e$ (\cmthree)               &$2.5\times 10^5$     & $ 2.5\times 10^5$   & $5\times 10^5$  \\
  size (pc)                      & 0.1                 &  0.1                & 0.05         \\
  N$_{HII}$                      & 1120                & 1.8$\times 10^{4}$   &  1280  \\
  N$_{HIIlos}$                   &1.7$\times 10^{-5}$  & 2.8$\times 10^{-4}$ & 4.8$\times 10^{-6}$ \\
  Area covering factor $f_c$     &1.7$\times 10^{-5}$  & 2.8$\times 10^{-4}$ & 4.8$\times 10^{-6}$ \\
  Volume filling factor $f_v$    &2.1$\times 10^{-9}$  & 3.4$\times 10^{-8}$ & 3.0$\times 10^{-10}$ \\
  $\tau_c$(8.3GHz)               & 34.8                & 23.8                & 47.5 \\
  $\tau_L$(8.3GHz)               & --7.5               & --4.3               & --8.5 \\
  $b_n$                          & 0.898               & 0.902               & 0.908\\
  $\beta_n$                      &--22.0               & --28.7              & --28.3         \\
  Stimulated emission (\%)       & 0.2                 &  0.2                &  0.2      \\
  $f_{nlte} $                    & 439                 &  35                 &  1282    \\
  S$_{thermal}$ (5GHz)(mJy)      & 0.01                &  0.2                &  0.004         \\
  S$_{non-th}$  (5GHz)(mJy)      & 200                 &  199                &  200      \\
  $\alpha_{non-th}$              &--0.60               & --0.61              & --0.60\\
 \mhii  (\msun)                  &3.6$\times 10^3$     &5.9$\times 10^4$     & 1.0$\times 10^3$    \\
  N$_{Lyc}$  (s$^{-1}$)          &3.5$\times 10^{53}$  &4.6$\times 10^{54}$  & 1.6$\times 10^{53}$ \\
  N$_{O5}$                       &7.6$\times 10^3$     &9.8$\times 10^4$     & 3.4$\times 10^3$    \\
  F$_{Br\alpha}$(W m$^{-2}$)&2.4$\times 10^{-17}$ &2.5$\times 10^{-16}$ & 8.7$\times 10^{-18}$\\
  F$_{Br\gamma}$(W m$^{-2}$)&8.5$\times 10^{-18}$ &9.2$\times 10^{-17}$ & 3.2$\times 10^{-18}$\\
  N$_{O5}$ per HII region        &     6.8             &      5.3             &      2.7\\
\hline
\end{tabular}
\end{table}

\begin{table}
{\bf Table 9} Parameters of the Three component   model \\

\begin {tabular}{lcccc}
\hline
Parameter &  Model A1 & Model A2 & Model D & Model A1+A2+D  \\
\hline

  T$_e$ (K)                 &  7500             & 7500                & 7500   & -- \\
  N$_e$ (\cmthree)         &$ 10^3$             & $2.5\times 10^5$    & 1000.  &--\\
  size (pc)                 &  5.0              & 0.1                 & 0.13   & -- \\
  EM (Pc \cmsix)            &  $5\times 10^6$   &$6.3\times 10^9$     &$1.3\times 10^5$& -- \\
  N$_{HII}$                 &  $1.9\times 10^4$ & 1120                &  --    & --\\
  N$_{HIIlos}$              &  0.7              & 1.7 $\times 10^{-5}$ &  --    & -- \\
  Area covering factor $f_c$&   0.7              & 1.7$\times 10^{-5}$ & 1.0 \\
  Volume filling factor $f_c$ & 4.3$\times 10^{-3}$& 2.1$\times 10^{-9}$&  --    &   \\
  $\tau_c$(8.3GHz)          & 0.03              & 34.8                & $7.2\times 10^{-4}$ & -- \\
  $\tau_L$(8.3GHz)          & --0.13            & --7.5               & $-5.5\times 10^{-5}$ & -- \\
  Stimulated emission (\%)  &  54               &  0.2                & --     & -- \\
  S$_{thermal}$ (5GHz)(mJy) &  31               &  0.01                &  1.2   & 32 \\
  S$_{non-th}$  (5GHz)(mJy) &  176              &  200                & --     & 176 \\
  $\alpha_{non-th}$         & --0.76            & --0.60              & --     & --0.8 \\
 \mhii  (\msun)             &3.0$\times 10^7$   & 3.6$\times 10^3$    & 2.1$\times 10^6$ & 3.2$\times 10^7$ \\
  N$_{Lyc}$  (s$^{-1}$)     &1.2$\times 10^{55}$& 3.5$\times 10^{53}$ & 8.3$\times 10^{53}$ & 1.3$\times 10^{55}$\\
  N$_{O5}$                  &2.5$\times 10^5$   & 7.6$\times 10^3$    & 1.8$\times 10^4$  & 2.8$\times 10^{5}$ \\
  F$_{Br\alpha}$(W m$^{-2}$) &9.2$\times 10^{-16}$& 2.4$\times 10^{-17}$ &6.5$\times 10^{-17}$&1.0$\times 10^{-15}$ \\
  F$_{Br\gamma}$(W m$^{-2}$) &2.9$\times 10^{-16}$ & 8.5$\times 10^{-18}$ & 2.0$\times 10^{-17}$&3.2$\times 10^{-16}$ \\
  N$_{O5}$ per HII region   &      13.6         &    5.3              & --\\    
\hline
\end{tabular}
\end{table}

\begin{table}
{\bf Table 10} Parameters from a Continuous Star Formation Model\tablenotemark{a} \\

\begin {tabular}{cccc}
\hline
$\tau_{csf}$ &  \nlyc                                    & \lbol                        & \lbol/$L_{L_\alpha}$  \\
$ 10^6$ yrs  & ph s$^{-1}$ /(\msun yr$^{-1}$) &  \lsun/(\msun yr$^{-1}$) &     (=IRE)                   \\
\hline

    0.5      &          9.2  $\times 10^{51}$         & 2.5 $\times 10^8$              & 6.3\\
    1.0      &         17.3 $\times 10^{51}$          & 5.0 $\times 10^8$              & 6.7\\
    5.0      &         36.9$\times 10^{51}$           &23.5 $\times 10^8$              & 14.9\\
   10.0      &         37.1$\times 10^{51}$           &41.0 $\times 10^8$              & 21.7\\
   15.0      &         37.1$\times 10^{51}$           &41.0$\times 10^8$               & 24.3\\
   20.0      &         37.1$\times 10^{51}$           &45.4$\times 10^8$               & 25.8\\
   50.0      &         37.1$\times 10^{51}$           &59.1$\times 10^8$               & 38.7\\
\hline
\end{tabular}
\tablenotetext{a}{Taken from Viallefond (1987)} 
\end{table}

\newpage
\begin{figure}[h]
\plotfiddle{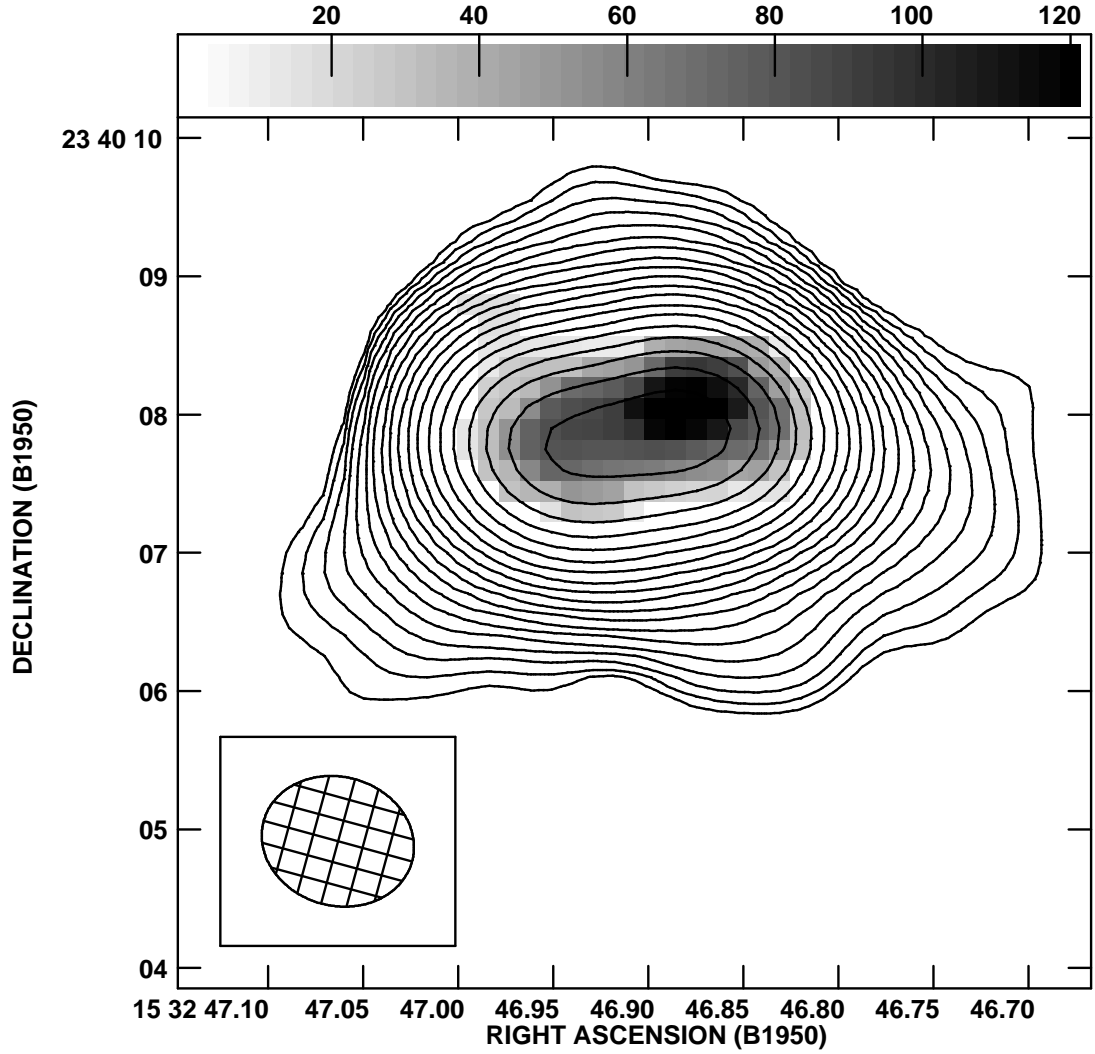}{ 6.5 in}{270}{75}{75}{-300}{500} 
\caption{
Continuum image of Arp 220 at 8.1 GHz  (contours),
made using the VLA,
with H92\a\- integrated line emission (moment 0) superposed (grey scale).
The beam shown at the bottom left corner is $1.1''\times 0.9''$, PA = 74\deg. The
peak continuum brightness is 82 mJy beam$^{-1}$. The first contour level
is 150 $ \mu$Jy beam$^{-1}$ and the contour levels progress geometrically
by a factor of 1.4. Grey scale for the integrated line intensity
covers the range 2 -- 120 Jy beam$^{-1}$ m s$^{-1}$.}
\end{figure}

\newpage
\begin{figure}[ht]
\plotfiddle{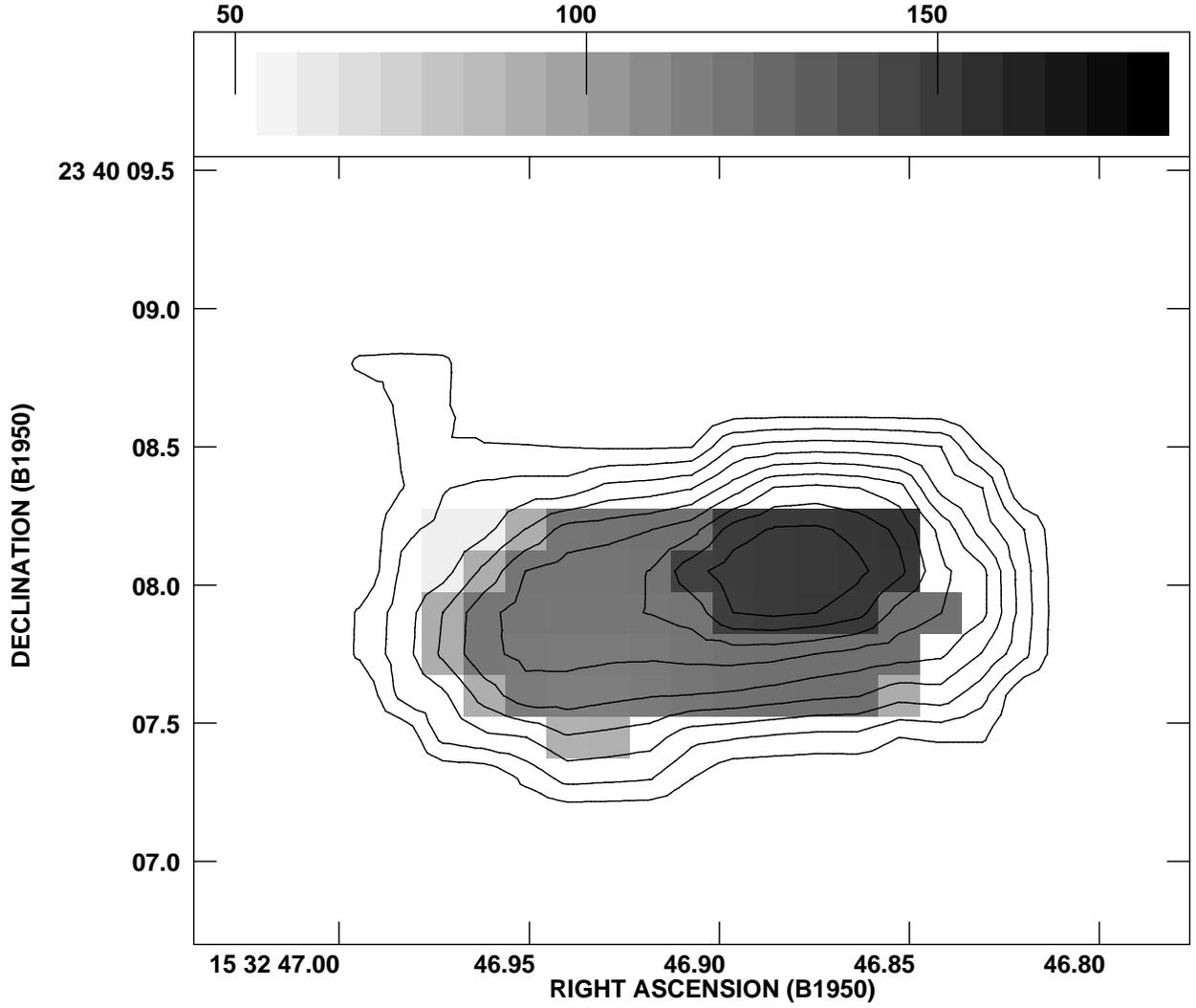}{ 6.5 in}{270}{75}{75}{-300}{500} 
\caption{
 Integrated H92\a\- line emission (moment 0) from Arp
220 in contours with velocity dispersion (moment 2) superposed in grey
scale.  Beam is same as in Fig 1. Contour peak flux is 130 Jy
beam$^{-1}$ m s$^{-1}$. Contour levels are 1, 2, 3 ... percent of the
peak. The grey scale covers the range 50 -- 180 \kms.}
\end{figure}

\newpage
\begin{figure}[ht]
\epsscale{0.6}
\plotfiddle{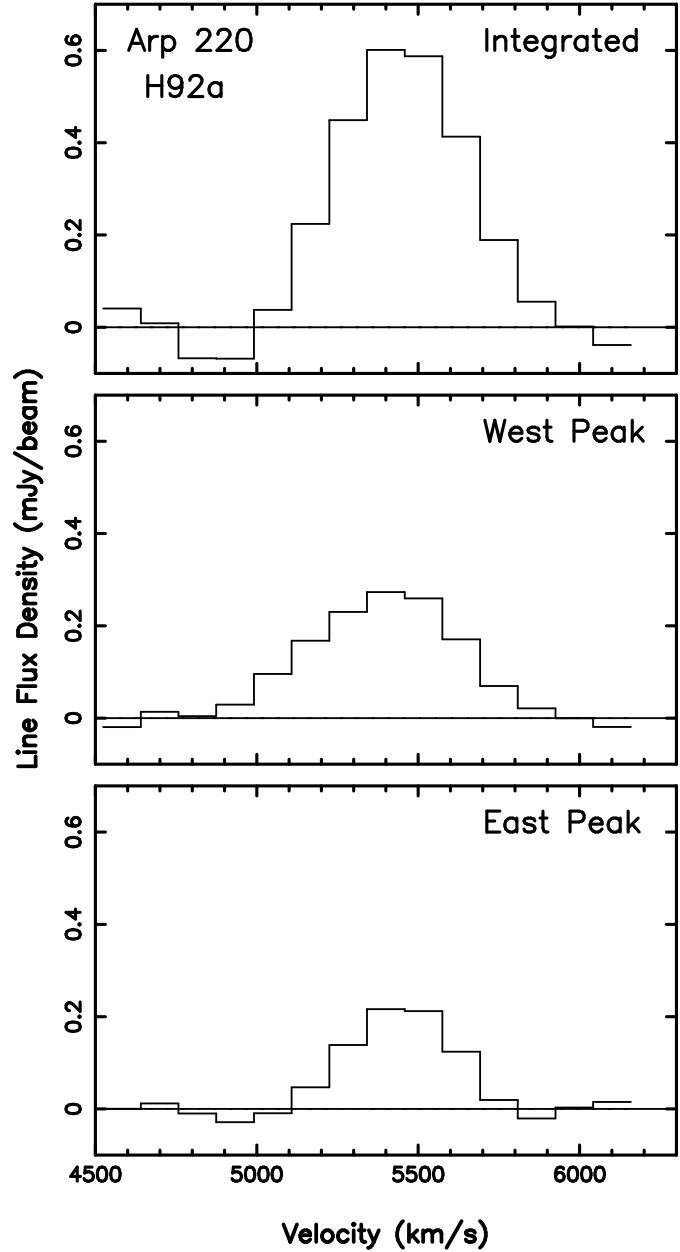}{ 5.0 in}{0}{65}{65}{-200}{-50} 
\vspace{0.5cm}
\caption{
H92\a\- line profiles in Arp 220 observed
using the VLA: {\em top:} Integrated
over the region shown as contours in Fig 2, {\em middle:} profile near
at the position RA (1950) = 15h32m46.87s, Dec (1950) = 23\deg 40'8.1'' near
the western peak and {\em bottom:} profile at the position RA (1950) = 
15h32m46.92s, Dec(1950) = 23\deg 40'7.8'' near the eastern peak. X-axis
is Heliocentric velocity.}
\end{figure}

\begin{figure}[ht]
\plottwo{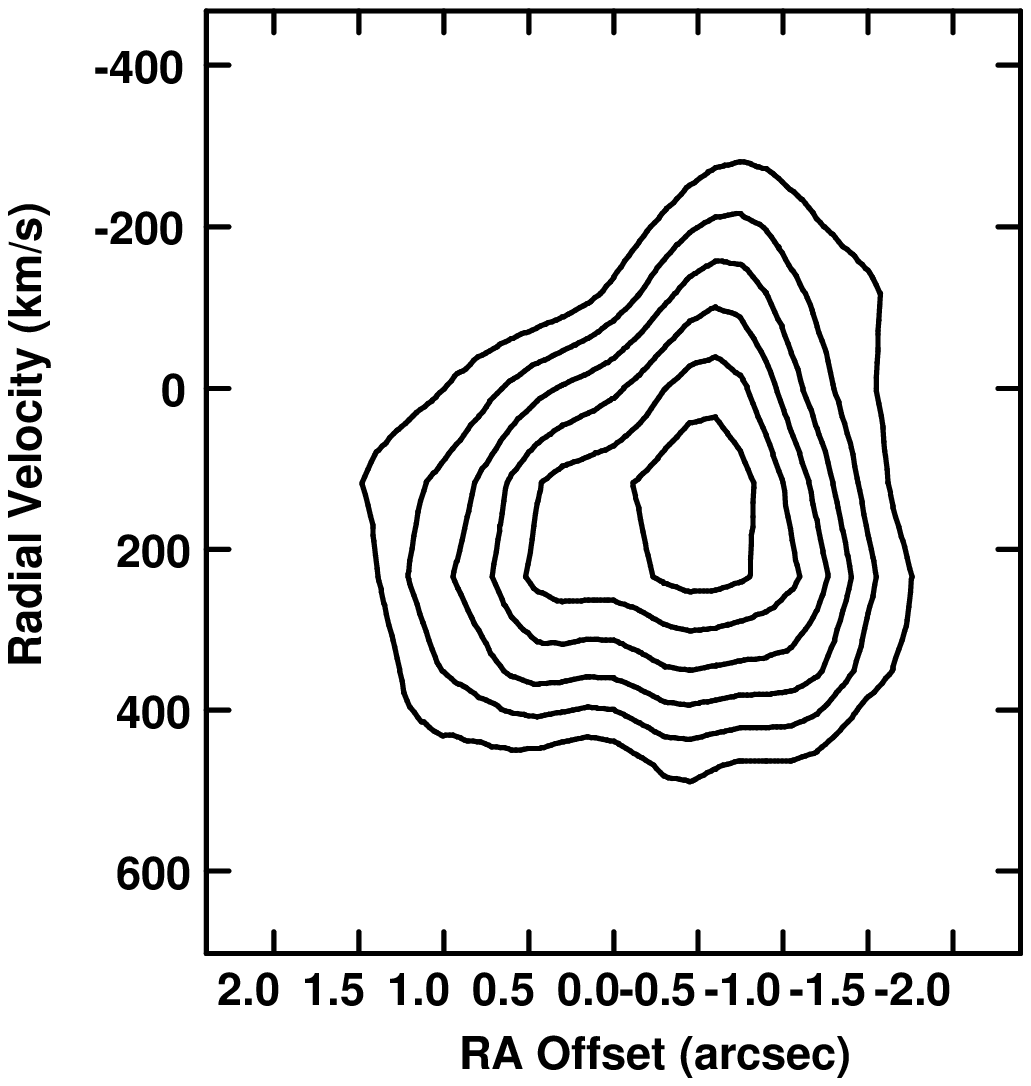}{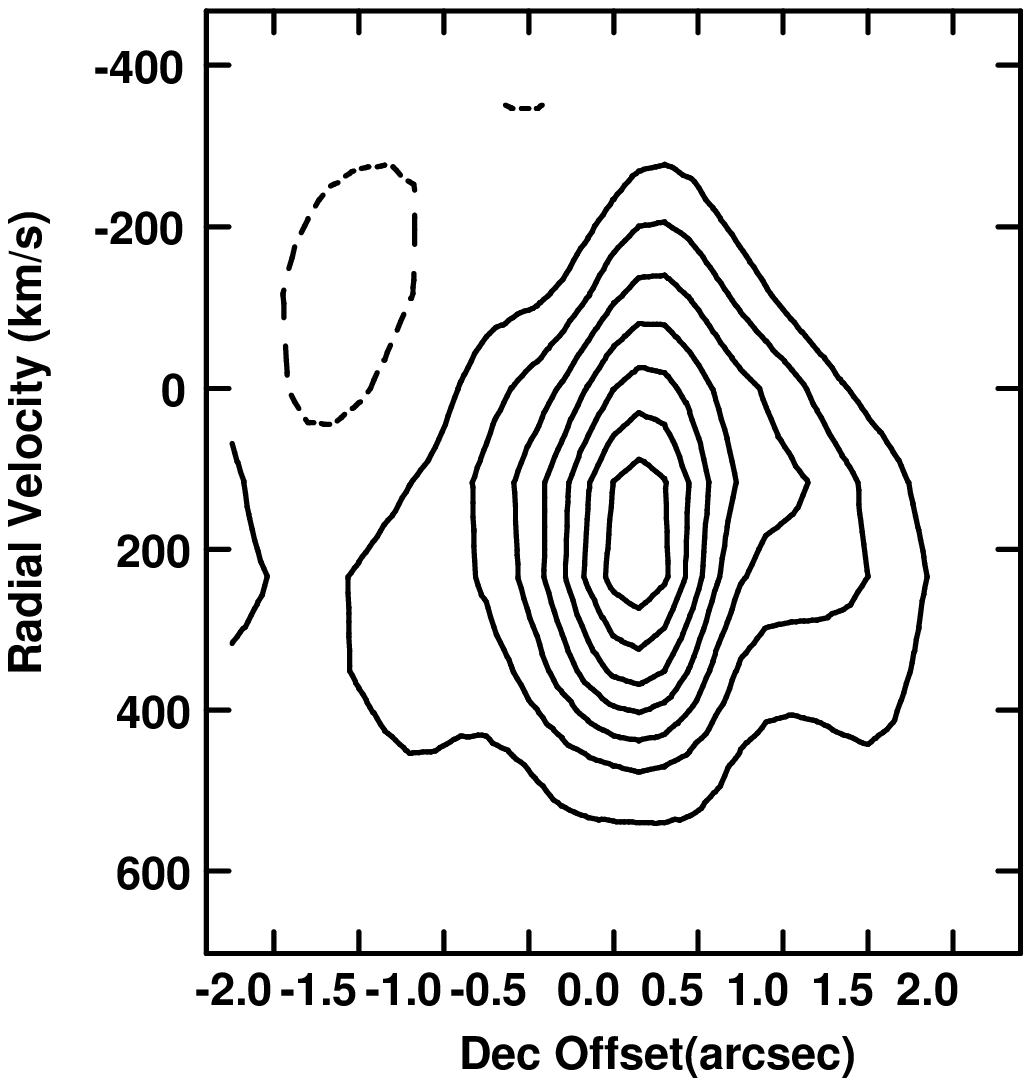}
\caption{(a)
Position-Velocity diagram of the H92\a- emission
along the line joining the
two continuum peaks in Arp 220 at a position angle of 98\deg. Offsets
along the X axis is from RA (1950) = 15h32m46.94s. Offsets along the
Y axis is from $V_{Hel}$ = 5283 \kms. The beam size is same as in
Fig 1. Peak brightness is 0.8 mJy/beam. Contour levels are 0.2 to 0.8 mJy/beam
at intervals of 0.1 mJy/beam.
(b) Position-Velocity diagram along a line perpendicular
to the line joining the two continuum peaks in Arp 220 at a position angle
of 8\deg. Offsets along the X axis is from DEC (1950) = 23\deg
40'7.5''. Offsets along the Y axis is from $V_{Hel}$ = 5283 \kms. The
beam size is same as in Fig 1. Peak brightness is 3.9 mJy/beam. Contour
levels are 0.5 to 4  mJy/beam at intervals of 0.5 mJy/beam.}
\end{figure}

\newpage
\begin{figure}[ht]
\epsscale{0.8}
\plotone{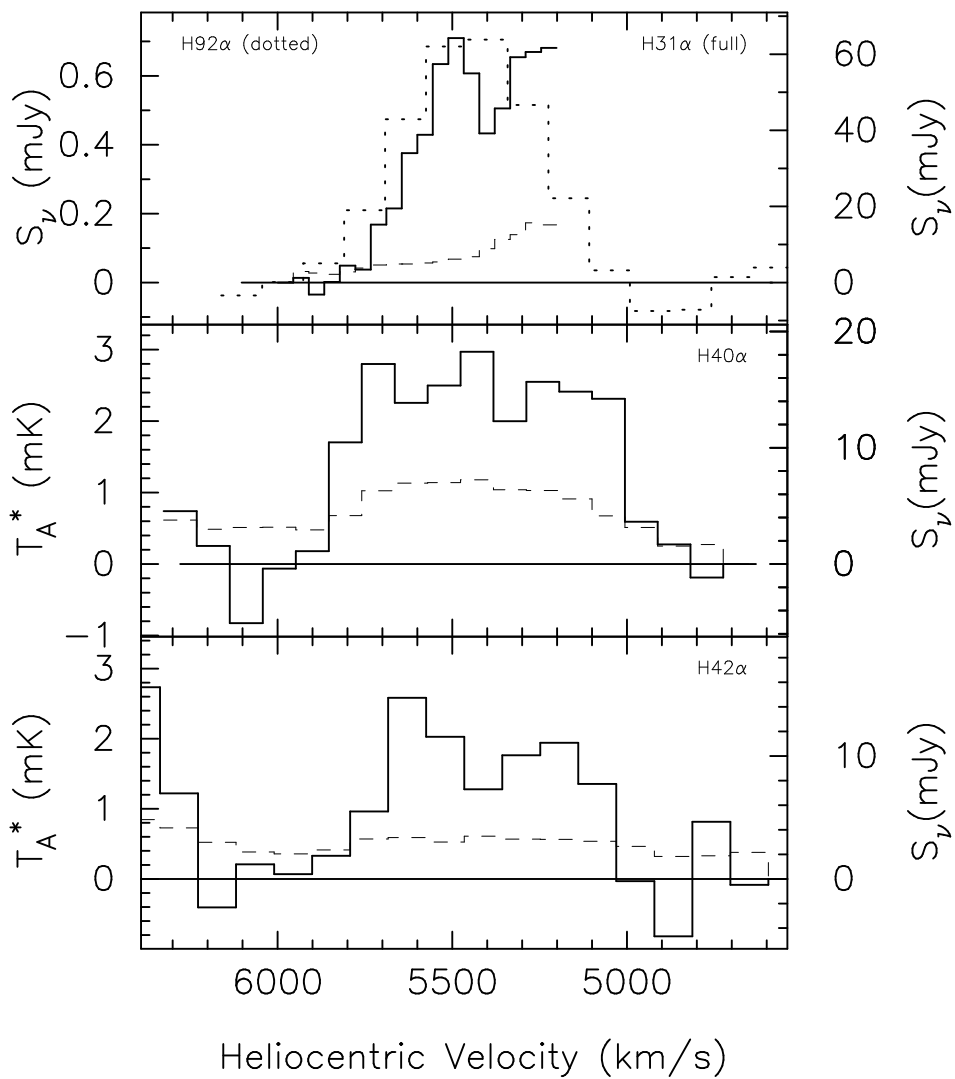}
\caption{
Recombination lines from Arp 220 in the 3mm (H40\a\- and H42\a) and
1.2 mm (H31\a) bands observed using the IRAM 30 m telescope. Solid
lines are the observed line profiles corrected for a linear
baseline. The dashed lines represent statistical rms noise level in
each channel including the uncertainty in the baseline level. The
dotted line in the top frame is the integrated H92\a\- line profile
from Fig 3.  For the middle and bottom frames, antenna temperature
corrected for atmospheric absorption is given on the left and the
equivalent flux density on the right. For the top frame, the scale on
the left hand side corresponds to the H92\a\- line and the one on the
right hand side to the H31\a\- line.  }
\end{figure}

\clearpage
\newpage
\begin{figure}[ht]
\plotone{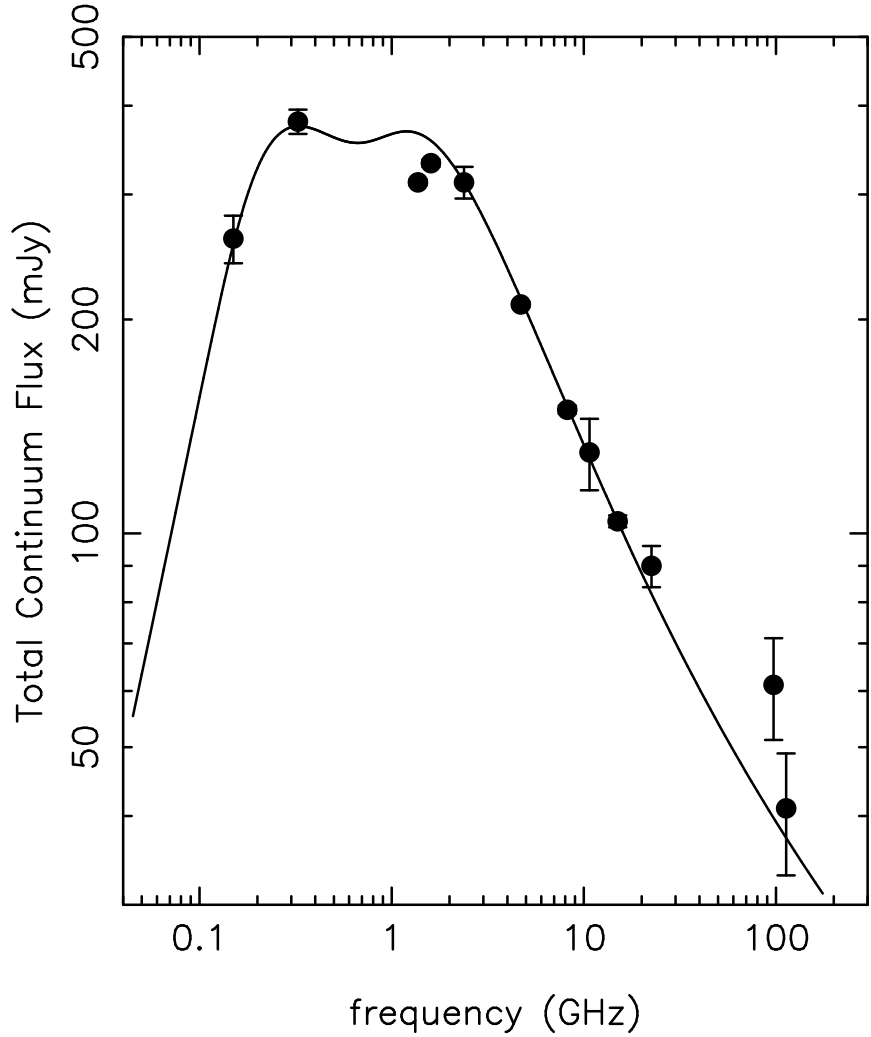}
\caption{
 Observed Continuum spectrum of Arp 220 over the
frequency range 0.15 GHz to 113 GHz. Data points are from Table 1.
The solid line is a model fit based on three ionized components and
one non-thermal component discussed in Section 4.2.3 (see also Fig
10). Formal error bar on each data point is also plotted. Data points
above 200 GHz are not plotted since they are dominated by dust
emission.}
\end{figure}

\newpage
\begin{figure}[ht]
\plotone{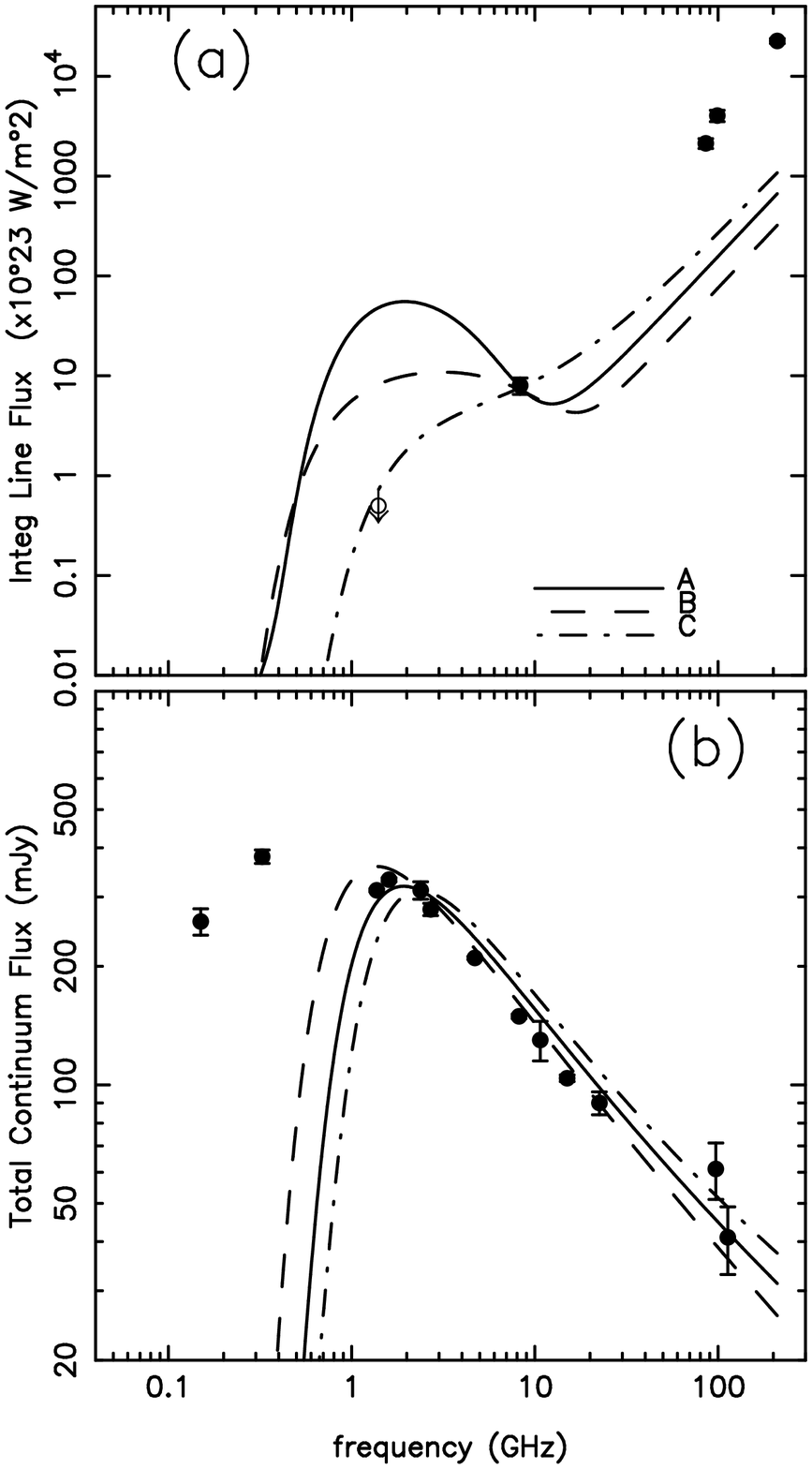}
\caption{
Expected variation of (a) recombination line and (b)
continuum strengths with frequency in Arp 220 based on uniform slab
models fitted to the H92\a\- recombination line near 8.3 GHz. The
parameters of the three models A, B and C are given in Table 6. The
observed continuum data points in (b) are from Table 5 and the
observed line data points in (a) are from Tables 2 and 4. Formal error bars
are plotted for all the data points. The line data point
near 1.4 GHz is an upper limit. These models are inconsistent with the
upper limit to the RRL at 1.4 GHz and the low frequency turnover in the
continuum spectrum. The models also cannot account for the high frequency
RRLs.}
\end{figure}

\newpage
\begin{figure}[ht]
\plotone{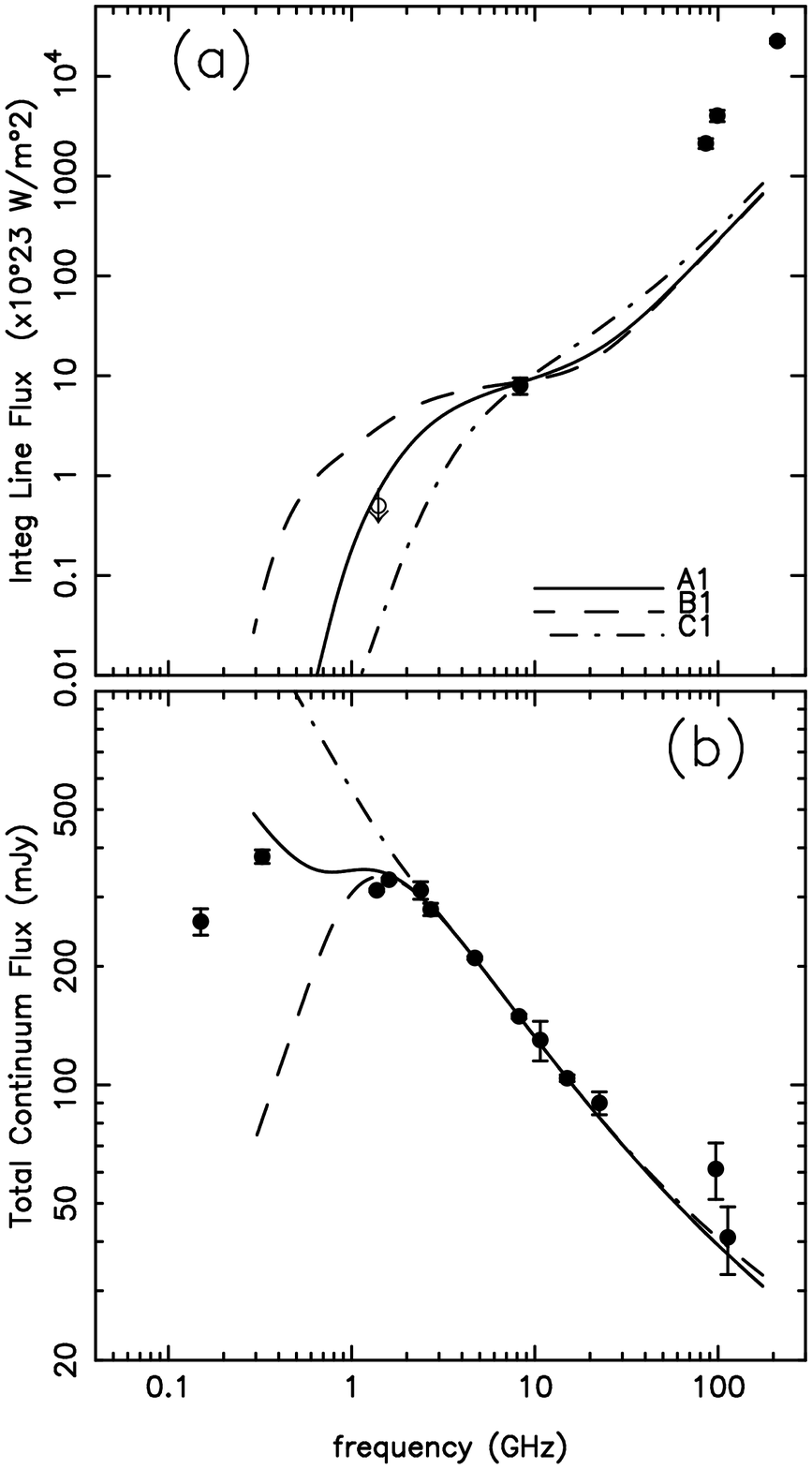}
\caption{
 Expected variation of (a) recombination line and (b)
continuum strengths with frequency in Arp 220 based on models with a
collection of HII regions fitted to the H92\a recombination line near
8.3 GHz. The parameters of the three models A1, B1 and C1 are given in
Table 7. The observed continuum data points in (b) are from Table 5
and the observed line data points in (a) are from Tables 2 and 4. Formal 
error bars are plotted for all the data points. The line
data point near 1.4 GHz is an upper limit. These models cannot explain the 
high frequency RRLs and the turnover in the continuum spectrum below
500 MHz.}
\end{figure}

\newpage
\begin{figure}[ht]
\plotone{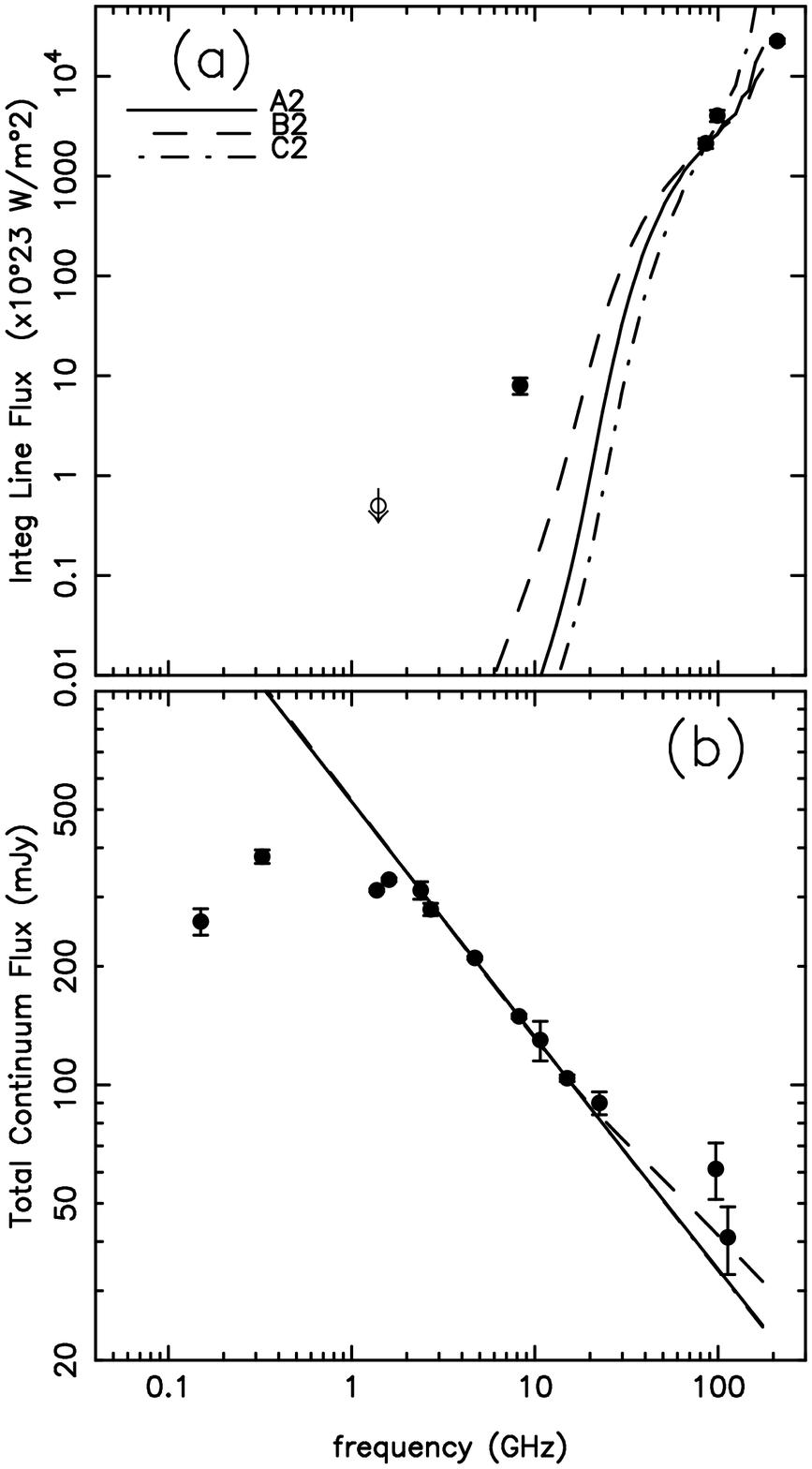}
\caption{
 Expected variation of (a) recombination line and (b)
continuum strengths with frequency in Arp 220 based on models with a
collection of HII regions fitted to the H42\a recombination line near
84 GHz. The parameters of the three models A2, B2 and C2 are given in
Table 8. The observed continuum data points in (b) are from Table 5
and the observed line data points in (a) are from Tables 2 and 4. Formal
error bars are plotted for all the data points. The line
data point near 1.4 GHz is an upper limit. These models can only explain the
high frequency RRLs.}
\end{figure}

\newpage
\begin{figure}[ht]
\plotone{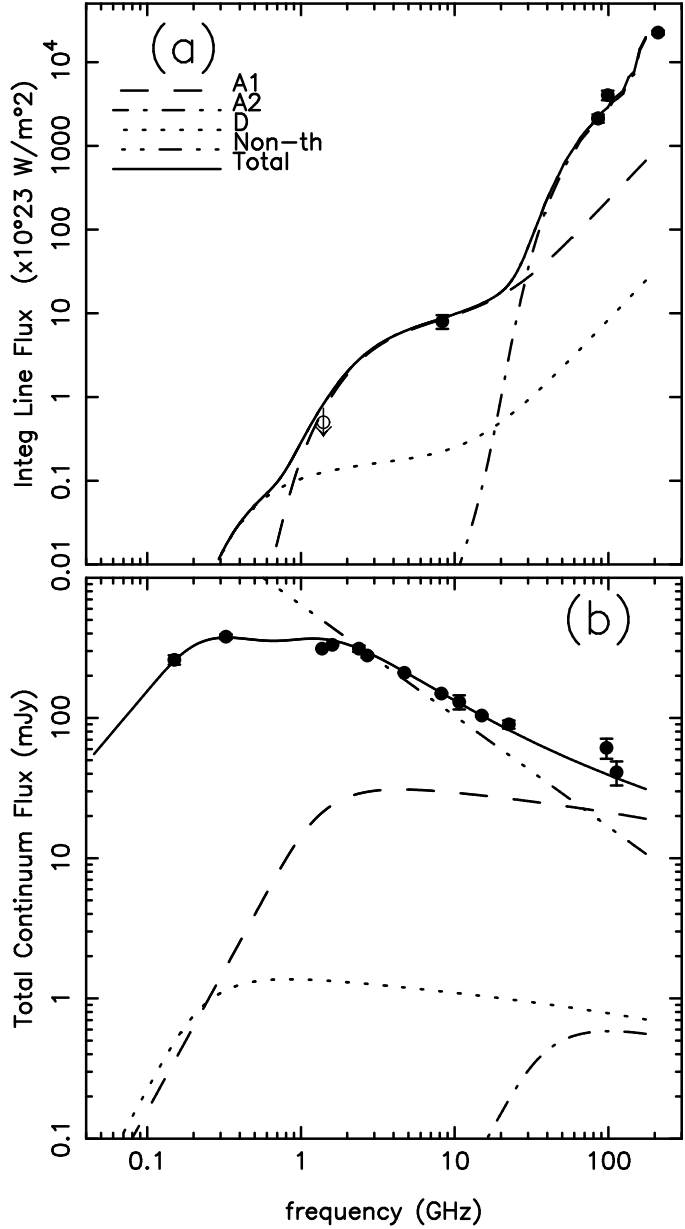}
\caption{
 Expected variation of (a) recombination line and (b)
continuum strengths with frequency in Arp 220 based on three ionized
components (A1, A2 and D) and one non-thermal component.  The
parameters of the three ionized components A1, A2 and D are given in
Table 9 (see also Tables 7 and 8). Contributions from the individual
components to line and continuum are shown as separate curves.  Keys
for the different curves are indicated at the top left corner of
(a). The non-thermal component shown in (b) has a spectral index of
--0.8 and a flux density of 176 mJy at 5 GHz. The observed continuum
data points in (b) are from Table 5 and the observed line data points
in (a) are from Tables 2 and 4. Formal error bars are also plotted for all
the data points. The line data point near 1.4 GHz is an
upper limit. The three component model is consistent with all the available
RRL and continuum observations.}
\end{figure}

\newpage
\begin{figure}[ht]
\plotfiddle{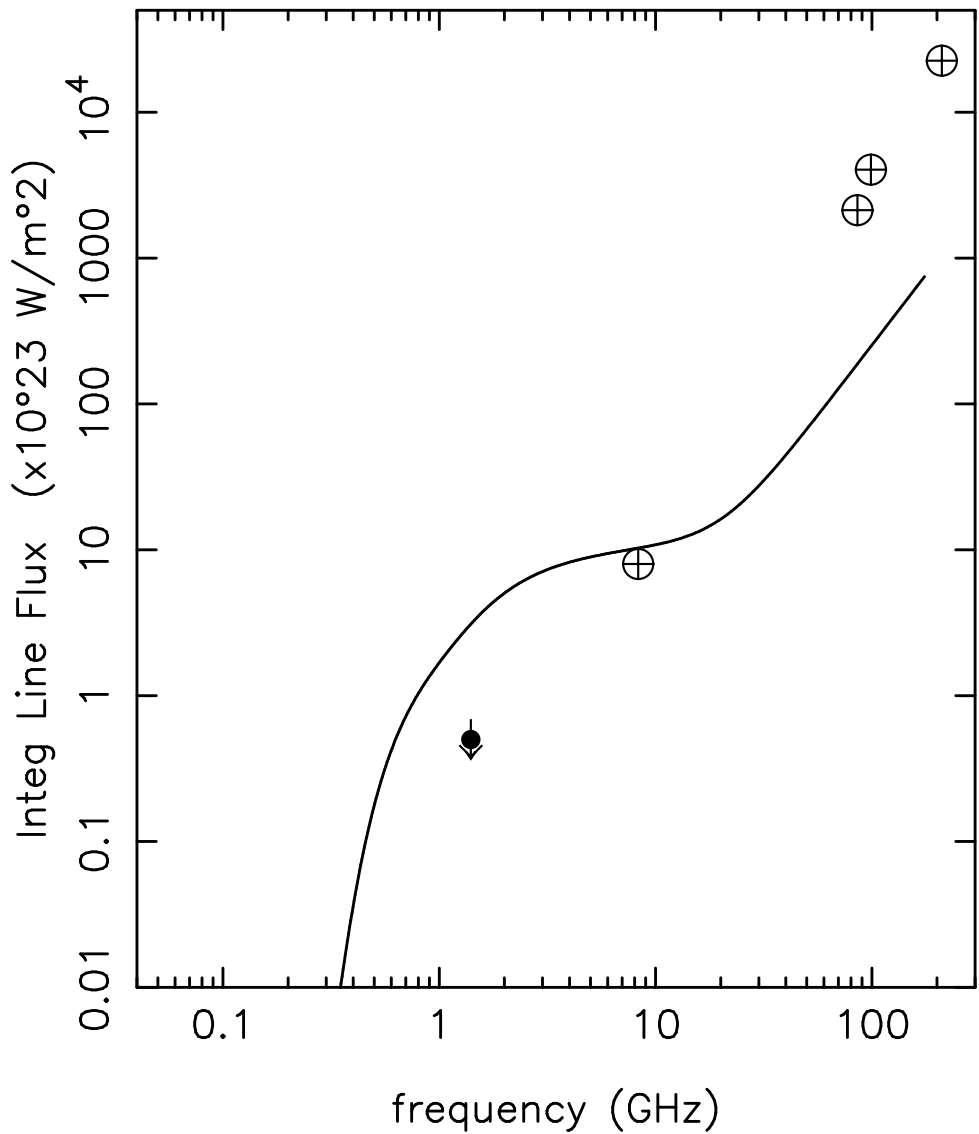}{ 6.5 in}{0}{75}{75}{-200}{-50} 
\caption{
Expected strength of radio recombination lines if the
density of ionized gas \ne = 500 \cmthree and production rate of Lyman
continuum photons \nlyc = $1.3 \times 10^{55}$ photons s$^{-1}$. These
parameters are similar to the values determined from NIR lines observed
with ISO. The observed line data points are from Tables 2 and 4. The
line data point near 1.4 GHz is a upper limit. For consistency with
the upper limit at 1.4 GHz, electron density must be $> 1000$
\cmthree. The observed recombination lines above 80 GHz require
ionized gas with \ne$> 10^5$ \cmthree.}
\end{figure}

\clearpage



\begin{thebibliography}{}
\bibitem[Anantharamaiah et al 1993]{} Anantharamaiah, K. R., Zhao,
        J. H., Goss, W. M., \&~Viallefond, F. 1993, ApJ, 419, 585 
\bibitem[Arp 1966]{} Arp, H. 1966, ApJS, 14, 1
\bibitem[Cornwell, Uson \& Haddad 1992]{} Cornwell, T. J., Uson, J. M.,
        \&~Haddad, N 1992, A\&A, 258, 583
\bibitem[Downes \&~Solomon 1998]{} Downes, D., \&~Solomon, P. M. 1998,
        ApJ, 507, 615
\bibitem[Elmegreen 1983]{} Elmegreen, B.G., 1983, MNRAS, 203, 1011
\bibitem[G\"{e}nzel et al 1998]{} G\"{e}nzel, R. et al 1998, ApJ, 498, 579
\bibitem[Goldader et al 1995]{} Goldader, J. D., Joseph, R. D., Doyon, R.,
         \& Sanders, D. B. 1995, ApJ, 444, 97
\bibitem[Graham et al 1990]{} Graham, J. R., Carico, D. P., Matthews, K.,
        Neugebauer, G., Soifer, B. T., \&~Wilson, T. D. 1990, ApJ, 354, L5
\bibitem[Houck et al 1984]{} Houck, J. R., Shure, M. A., Gull, G. E., \&
         Herter, T. 1984, ApJ, 287, L11
\bibitem[Kennicutt 1998]{} Kennicutt, R. C., JR. 1998, ApJ, 498, 541
\bibitem[Lonsdale et al 1998]{} Lonsdale, C. J., Lonsdale, C. J., Diamond,
         P. J., \& Smith, H. E. 1998, ApJ, 493, L13
\bibitem[Lutz et al 1996]{} Lutz, D. et al 1996, A\&A, 315, L137
\bibitem[Mathis 1986]{} Mathis, J. S. 1986, PASP, 98, 995
\bibitem[Mezger 1985]{} Mezger, P. G. 1985 in Birth and Infancy of
         Stars(Amsterdam, North-Holland), 31 
\bibitem[Mezger 1986]{} Mezger, P. G. 1986, Ap\&SS, 128, 111
\bibitem[Mezger et al 1974]{} Mezger, P. G., Smith, L. F., \& Churchwell,
         E. 1974, A\&A, 32, 269
\bibitem[Miller \&~Scalo 1978]{} Miller, G. E., \&~Scalo, J. M. 1978,
         PASP, 90, 506
\bibitem[Myers et al 1986]{} Myers, P.C., Dame, T.M., Thaddeus, P., Cohen,
        R.S., Silverberg, R.F., Dwek, E. \& Hauser, M.G. 1986, ApJ, 301, 398
\bibitem[Norris 1988]{} Norris, R. P. 1988, MNRAS, 230, 345
\bibitem[Panagia 1978]{} Panagia, N. 1978 in Infrared Astronomy,
         Proceedings of the Advanced Study Institute(Dordrecht, D. Reidel
         Publishing Co.), 115
\bibitem[Phookun et al 1998]{} Phookun, B., Anantharamaiah, K. R., \&
         Goss, W. M. 1998, MNRAS, 295, 156
\bibitem[Puxley et al 1991]{} Puxley, P. J., Brand, P. W. J. L., Moore, T.
         J. T., Mountain, C. M., \& Nakai, N. 1991, MNRAS, 248, 585
\bibitem[Rice et al 1990]{} Rice, W., Boulanger, F., Viallefond, F.,
         Soifer, B. T., \& Freedman, W. L. 1990, ApJ, 358, 418
\bibitem[Rigopoulou et al 1996]{} Rigopoulou, D., Lawrence, A.,
         Rowan-Robinson, M. 1996, A\&A, 305, 747
\bibitem[Sakamoto et al 1999]{} Sakamoto, K., Scoville, N. Z., Yun, M. S.,
         Crosas, M., G\"{e}nzel, R., Tacconi, L. J. 1999, ApJ, 514, 68
\bibitem[Salpeter 1955]{} Salpeter, E. E. 1955, ApJ, 121, 161
\bibitem[Sanders \&~Mirabel 1996]{} Sanders, D. B., Mirabel, I. F. 1996,
         ARA\&A, 34, 749
\bibitem[Scoville et al 1998]{} Scoville, N. Z. et al 1998, ApJ, 492, L107
\bibitem[Scoville et al 1997]{} Scoville, N. Z., Yun, M. S., \&~Bryant, P.
         M. 1997, ApJ, 484, 702
\bibitem[Bell \&~Seaquist 1977]{} Bell, M. B., \&~Seaquist, E. R. 1977,
         ApJ, 223, 378 
\bibitem[Shaver et al 1977]{} Shaver, P. A., Churchwell, E., \&~Rots, A.
         H. 1977, A\&A, 55, 435
\bibitem[Silk 1977]{} Silk, J. 1977, ApJ, 214, 718
\bibitem[Smith et al 1998]{} Smith, H. E., Lonsdale, C. J., Lonsdale, C.
         J., \&~Diamond, P. J. 1998, ApJ, 493, L17 
\bibitem[Solomon et al 1990]{} Solomon, P. M., Radford, S. J. E.,
          \&~Downes, D. 1990, ApJ, 348, L53
\bibitem[Sopp \&~Alexander 1991]{} Sopp, H. M., \&~Alexander, P. 1991,
        MNRAS, 251, 14
\bibitem[Spitzer 1978]{} Spitzer, L. 1978, Physical Processes in the
        Interstellar Medium(New York Wiley-Interscience)
\bibitem[Sturm et al 1996]{} Sturm, E. et al 1996, A\&A, 315, L133
\bibitem[Viallefond et al 1982]{} Viallefond, F., Goss, W. M., \&~Allen,
        R. J. 1982, A\&A, 115, 373
\bibitem[Viallefond 1987]{} Viallefond, F. 1987, These d'Etat, Universite Paris VII.
\bibitem[Zhao et al 1996]{} Zhao, J. H., Anantharamaiah, K. R., Goss, W.
        M., \&~Viallefond, F. 1996, ApJ, 472, 54
\bibitem[van der Hulst at al 1992]{} van der Hulst, J. M., Terlouw, J. P.,
        Begeman, K. G., Zwitser, W., \&~Roelfsema, P. R. 1992, in ASP Conf. Ser.
        25, Astronomical Data Analysis Software and Systems I, ed. D. M. Worrall,
        C. Biemesderfer, \&~J. Barnes, 131
\end{thebibliography}
\end{document}